\documentclass[twocolumn,preprintnumbers,superscriptaddress,amsmath,amssymb,prb]{revtex4-2}

\usepackage{booktabs} 
\usepackage{multirow} 
\usepackage{amsmath}  

\usepackage{graphicx}
\usepackage{dcolumn}
\usepackage{subfigure}
\usepackage{bm}
\usepackage[utf8]{inputenc}  
\usepackage{comment}
\usepackage{units}
\usepackage[unicode=true, bookmarks=false, breaklinks=false, pdfborder={0 0 1},colorlinks=false]{hyperref}
\usepackage{breakurl}
\usepackage{url}
\usepackage{natbib}
\usepackage{multirow}
\usepackage{float}
\usepackage{xcolor}

\renewenvironment{widetext@grid}{%
  \par\ignorespaces
  \setbox\widetext@top\vbox{%
   \vskip15\p@
   \hb@xt@\hsize{%
    \leaders\hrule\hfil
    \vrule\@height6\p@
   }%
   \vskip6\p@
  }%
  \setbox\widetext@bot\hb@xt@\hsize{%
    \vrule\@depth6\p@
    \leaders\hrule\hfil
  }%
  \onecolumngrid
  \let\set@footnotewidth\set@footnotewidth@ii
}{%
  \par
  \twocolumngrid\global\@ignoretrue
  \@endpetrue
}%
\renewcommand{\vec}[1]{\boldsymbol{#1}}

\begin{document}

\title{Chemical disorder effects on the Gilbert damping of FeCo alloys} 

\newcommand{\WISEKTH}{Wallenberg Initiative Materials Science for Sustainability (WISE), KTH Royal Institute of Technology, SE-10044 Stockholm, Sweden}
\newcommand{\WISEUU}{Wallenberg Initiative Materials Science for Sustainability (WISE), Uppsala University, Box 516, SE-75120 Uppsala, Sweden}
\newcommand{\KTH}{Department of Applied Physics, School of Engineering Sciences, KTH Royal Institute of Technology, 
AlbaNova University Center, SE-10691 Stockholm, Sweden}
\newcommand{\SeRC}{SeRC (Swedish e-Science Research Center), KTH Royal Institute of Technology, SE-10044 Stockholm, Sweden}
\newcommand{\Uppsala}{Department of Physics and Astronomy, Uppsala University, Box 516, SE-75120 Uppsala, Sweden}
\newcommand{\Orebro}{School of Science and Technology, \"Orebro University, SE-701 82, \"Orebro, Sweden}
\newcommand{\Stockholm}{Department of Materials and Environmental Chemistry, Stockholm University, SE-10691 Stockholm, Sweden}

\author{Zhiwei Lu}
\thanks{These two authors contributed equally}
    \affiliation{\KTH}
    \email[Corresponding author:\ ]{zhiweil@kth.se}
\author{I. P. Miranda}
\thanks{These two authors contributed equally}
    \affiliation{\Uppsala}
    \email[Corresponding author:\ ]{ivan.miranda@alumni.usp.br}
\author{Simon Streib}
    \affiliation{\Uppsala}
    
\author{Qichen Xu}
    \affiliation{\KTH}
\author{Rajgowrav Cheenikundil}
 \affiliation{\Orebro}

\author{Manuel Pereiro}
    \affiliation{\Uppsala}

\author{Erik Sjöqvist}
    \affiliation{\Uppsala}
    
\author{Olle Eriksson}
    \affiliation{\Uppsala}
    \affiliation{\Orebro}
    \affiliation{\WISEUU}
    
\author{Anders Bergman}
    \affiliation{\Uppsala}
    
\author{Danny Thonig}
    \affiliation{\Orebro}
    \affiliation{\Uppsala}

\author{Anna Delin}
    \affiliation{\KTH}
    \affiliation{\SeRC}
    \affiliation{\WISEKTH}
\date{\today}

\begin{abstract}
The impact of the local chemical environment on the Gilbert damping in the binary alloy Fe$_{100-x}$Co$_{x}$ is investigated, using computations based on density functional theory.
By varying the alloy composition $x$ as well as Fe/Co atom positions we reveal that the effective damping of the alloy is highly sensitive to the nearest neighbor environment, especially to the amount of Co and the average distance between Co-Co atoms at nearest neighbor sites. Both lead to a significant local increase (up to an order of magnitude) of the effective Gilbert damping, originating mainly from variations of the density of states at the Fermi energy. In a global perspective (i.e., making a configuration average for a real material), those differences in damping are masked by statistical averages.
When low-temperature explicit atomistic dynamics simulations are performed, the impact of short-range disorder on local dynamics is observed to also alter the overall relaxation rate. Our results illustrate the possibility of local chemical engineering of the Gilbert damping, which may stimulate the study of new ways to tune and control materials aiming for spintronics applications.
\end{abstract}

\maketitle 

\section{Introduction}

Magnetic damping, also known as Gilbert damping ($\alpha$) in the context of the Landau-Lifshitz-Gilbert (LLG) equation \cite{eriksson2017atomistic}, plays a crucial role in determining the rate at which energy and angular momentum dissipate in a magnetic system. It thus generally dictates the timescale of processes involving a magnetic transition from an excited state to an equilibrium state, magnetic domain wall motion, and spin-wave propagation. These are the fundamental processes that govern the performance of spintronics \cite{chumak2015magnon} and magnonics \cite{barman20212021} devices. Reduced magnetic damping can lead to higher efficiency and faster operations in these devices by minimizing energy dissipation during their operation. Hence, a detailed investigation into the fundamental physics involved in the mechanism of damping is of great interest \cite{schoen2016ultra, jungwirth2016antiferromagnetic, chumak2015magnon, yang2022two}. 
Previous theoretical and experimental research has established that $\alpha$ is not a simple scalar quantity (as is usually assumed), but instead a tensorial quantity \cite{PhysRevB.72.064450, PhysRevB.78.020410, Mondal2018, PhysRevB.98.104406, PhysRevLett.108.057204, PhysRevLett.102.086601} that is temperature-dependent \cite{Zhao2016, PhysRevB.87.014430, Hiramatsu2021}, non-local in real-space \cite{thonig2014gilbert, PhysRevMaterials.2.013801, PhysRevB.72.064450, PhysRevB.78.020410, PhysRevB.105.155151, PhysRevB.108.014433, PhysRevLett.113.237204, Weissenhofer2024, Reyesosorio2023}, and presents an anisotropic behavior -- as outlined in various studies  \cite{PhysRevMaterials.2.013801, PhysRevB.103.L220405, PhysRevLett.122.117203, brinker2022generalization, PhysRevB.104.024404}.
Therefore, fundamental understanding and materials engineering at the local (atomistic) scale compose, in principle, a possible route to optimum properties with desired damping magnitude.

Apart from their exceptional properties, including high Curie temperature, high saturation magnetization and high permeability \cite{sundar2005soft}, Fe-Co alloys constitute also a suitable platform to the fundamental investigation of the magnetic damping. It was experimentally demonstrated that the Fe$_{100-x}$Co$_x$ alloys (up to $x=60\%$) can hold an ultra-low intrinsic damping, of the order of $10^{-4}$, at $x\sim25\%$ \cite{schoen2016ultra}, despite being metallic. This surprising result, comparable to the values found in ferrimagnetic insulators (such as yttrium iron garnet \cite{Kajiwara2010}), was explained by the existence, at this particular value of $x$, of a sharp minimum density of states at the Fermi level, and corroborated by different methods and levels of theory \cite{PhysRevLett.107.066603, PhysRevB.87.014430, PhysRevB.98.104406, PhysRevB.108.014433, PhysRevB.98.174412, PhysRevB.92.214407}. In addition, Fe-Co has recently attracted attention with the observation of a giant anisotropy effect in $\alpha$ \cite{PhysRevLett.122.117203} in Fe$_{50}$Co$_{50}$ thin films, initially attributed to the variation of spin-orbit coupling due to local distortions in the alloy. This was also experimentally verified for other similar alloys involving Fe and Co \cite{PhysRevApplied.19.024030,Wang2023}, but the exact mechanisms responsible for this effect is still an open question \cite{PhysRevB.103.L220405,PhysRevB.103.L220403,PhysRevApplied.19.024030,PhysRevLett.131.186703}. Additionally, at the theoretical level, Lu \textit{et al.} \cite{PhysRevB.108.014433} suggested that for $x\in\{30\%,50\%\}$ the explicit treatment of damping as a nonlocal parameter in spin-dynamics have the effect of increased magnon lifetime predictions (for specific wave vectors $\mathbf{q}$). Together, those observations clearly consolidate FeCo as a benchmark for Gilbert damping investigations, encompassing many hidden features across the alloy series.

Despite the progress in studying these interesting phenomena in FeCo, other potential sources of hidden features are still not quite well-explored in the scope of Gilbert damping. Some have been recently highlighted as effects that are or might be relevant, also accounting for site-nonlocal damping  \cite{PhysRevB.109.094417,PhysRevB.103.L220405}. They are often related to changes in the local environment: reduced dimensionality, chemical disorder, and thermal fluctuations in the lattice (atomic displacements) and in the magnetic state. Although the former remains as the least discussed among the known effects \cite{PhysRevB.109.094417}, the other two, which constitute causes for electron scattering processes, are generally addressed by effective alloy analogy models (see, e.g., Refs. \cite{PhysRevB.91.165132,PhysRevLett.107.066603}). In particular, an explicit atomistic picture is still missing, and here we choose to address this knowledge gap in the context of the \textit{chemical disorder}. 

In this sense, the alloy models generally used -- the virtual crystal approximation \cite{PhysRevB.45.12911} or the coherent potential (CPA) approximation \cite{faulkner1982modern} -- were originally developed to describe chemical disorder in materials \cite{PhysRevLett.52.77} via an effective medium. The effective medium is a result of locally averaging possible clusters with different configurations and compositions within the alloy, and the CPA is designed to give the scattering properties of electron states that correspond to a proper configuration average of the alloy. This theoretical description 
works reasonably well to reproduce experimentally measured electronic structure properties of many alloys \cite{PhysRevLett.93.027203,PhysRevB.36.9657,PhysRevLett.40.900,PhysRevB.66.214206} and related materials properties, including Gilbert damping \cite{PhysRevB.87.014430,PhysRevB.92.214407,PhysRevB.94.214410,PhysRevB.108.014433}. Nevertheless, they frequently overlook the impacts of the local environment, including atom clustering, among others -- precisely the conditions under which phenomena like damping anisotropy were originally suggested to arise \cite{PhysRevLett.122.117203}. 

The local chemical environment has been shown to modify the magnetic properties of a given material, such as the atomic magnetic moment, magnetic exchange interactions \cite{bezerra2013complex, miranda2017dimensionality, de2016first, martins2016magnetic}, as well as  magnetic anisotropy \cite{PhysRevB.97.224402}. For instance, the presence of heterogeneous easy-axis orientations in clusters leads to a vanishing low magnetic anisotropy in permalloy (Ni$_{80}$Fe$_{20}$), although the favored easy-axis orientations at each site, dependent on the local atomic arrangement and the cluster's composition, can be several times larger than that of elemental Fe or Ni \cite{PhysRevB.97.224402}. As a result, the spin Hamiltonian and, thus, the energy landscape is modified, impacting the dynamics of the local magnetic moments in a system described by the LLG equation \cite{PhysRevB.86.224401}. In the context of FeCo, early calculations of ordered bcc Fe$_{50}$Co$_{50}$ structures have demonstrated that the magnetic anisotropy can change by some two orders of magnitude when particular Fe/Co configurations are considered \cite{Razee2001}. Also, experimental studies have shown that the magnetic properties are influenced by various local environment factors, including composition, morphology \cite{hesani2010effect, PhysRevLett.96.257205, karipoth2016magnetic}, as well as the size of Fe-Co alloy nanoclusters. 

In this study, we utilize Fe-Co alloys as a platform to investigate the impact of local compositions and atom arrangements on $\alpha$ via \textit{ab-initio} calculations. Here, we suggest the existence of hidden features (e.g., enhanced or decreased damping values) when an explicit atomistic picture is considered. As a consequence, local chemical engineering may be considered as a possible route to manipulate the Gilbert damping of such materials. 

The paper is organized as follows. In Sec. \ref{sec:methods}, we provide details on the density functional theory (DFT) calculations, the embedded cluster virtual crystal approximation model, along with the considered atomic configurations with different Co concentrations. Section \ref{sec:results} presents the results of the local environment effects on the Gilbert damping. In Sec. \ref{sec:conclusion}, we give a summary and an outlook.
\section{Method}
\label{sec:methods}

\subsection{Computational details}
\label{subsec:DFT}

The electronic structure and, from it, the magnetic properties of the Fe-Co alloys were obtained using the fully self-consistent real-space linear muffin-tin orbital in the atomic sphere approximation (RS-LMTO-ASA) \cite{PhysRevB.44.13283,PhysRevB.46.14570}. In this first-principles scheme, the eigenvalue problem is solved directly in the real-space by employing Haydock recursion method \cite{Haydock1980} (for $LL$ recursion steps) together with Beer-Pettifor terminator \cite{Beer1984}, suitable for metallic systems. Here, the local spin density approximation (LSDA), with parametrization by von Barth and Hedin \cite{Barth1972}, was used for the exchange-correlation functional (XC). To evaluate the Gilbert damping in the torque-correlation method, the spin-orbit coupling (SOC) term, $\xi\hat{L}\cdot\hat{S}$, was included self-consistently at each variational step \cite{PhysRevB.69.104401}. 
The LMTO-ASA \cite{PhysRevB.12.3060} is a linear method that gives precise results around a given energy $E=E_{\nu}$, often chosen as the center of gravity of the $s$, $p$ and $d$ bands. Thus, as in the previous study \cite{PhysRevB.108.014433}, we consider an expression accurate to $(E-E_{\nu})^2$ starting from the orthogonal representation of LMTO-ASA -- given that nonlocal damping parameters are fine quantities. For all cases investigated, we considered $LL=31$, which produces reliable results for the damping of Fe-Co alloys when compared to available experimental and theoretical results in the literature \cite{PhysRevB.108.014433}.

In order to simulate alloys, we designed the following simulation protocol: we performed \textit{ab-initio} calculations on a virtual-crystal-approximation (VCA) medium of bcc FeCo alloys to account for the chemical disorder. In general, this approximation is well-known to be reasonable for systems where the energy bands behave more or less rigidly with changes in the alloy concentration -- which is exactly the case of FeCo, neighboring elements in the periodic table (for a more detailed discussion on the VCA for FeCo, see Appendix \ref{VCA}). Here, the medium matrix contains $\sim55000$ atoms. The Fe-Co alloys still maintain a body-centred-cubic (bcc) phase for the range of Co concentration from 0\% to 60\% \cite{schoen2016ultra} and the lattice parameter $a_0=2.87$\AA$\,$ was used for all values of $x$. Although the FeCo alloys present a slight variation of $\sim1\%$ in $a_0$ for $x\in[0\%,60\%]$ \cite{PhysRevB.93.224425,Ohnuma2002}, we keep $a_0$ fixed to concentrate on the effect coming solely from the chemical disorder. It also should be noticed that the composition of $x=60\%$ Co is at the edge of the transition to body-centred tetragonal (bct) phase \cite{andersson2006structure}, but for simplicity, we remain here in the bcc phase. As for the VCA approach we replace the nuclear charge ($Z$) and the number of valence electrons ($n_v$) by the alloy average based on the concentration, e.g., in the case of Fe$_{60}$Co$_{40}$ we consider $n_v=8.4$ and $Z=26.4$.

\begin{figure}[!ht] 
\centering
        \includegraphics[width=0.7\columnwidth]{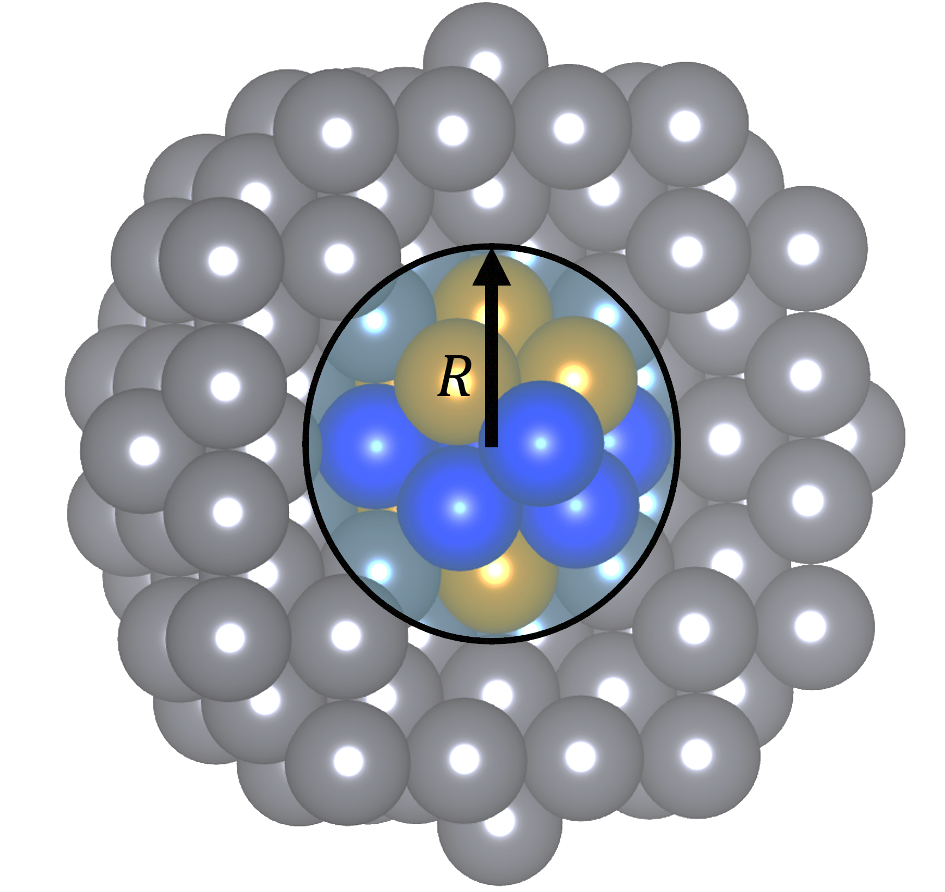}
        \caption{(Color online) Schematic representation of the EC-VCA model in the bcc structure. The Fe and Co atoms are depicted by yellow and blue spheres, respectively, forming an explicit Fe$_{53}$Co$_{47}$ embedded cluster with radius $R$. In turn, the gray spheres symbolize the VCA medium mimicking the same alloy concentration, that extends significantly beyond the depicted limits (the full VCA matrix has $\sim55000$ atoms). Image produced with the VESTA software \cite{Momma2011}.} 
        \label{fig:ec-vca-representation} 
\end{figure}

After the self-consistent procedure on the pure VCA structure, $15$-atom clusters composed of different Fe and Co compositions as well as atomic arrangements were embedded into the center of the corresponding VCA medium as local impurities (see Figure \ref{fig:ec-vca-representation} for a schematic representation).

The impurity region was self-consistently treated while the potential parameters and Fermi energy of the pure VCA matrix host were kept fixed. This is a particularly suitable approach and certainly an improvement with respect to the simple VCA approximation, because the short-range disorder (which pure VCA neglects) is made explicit, while the more distant sites are replaced by the virtual atoms with long-range disorder properties. Naturally, this model tends to become more accurate whenever the radius $R$ of the explicit region increases, but with two main limitations: (\textit{i}) $R$ cannot be too close from the full medium matrix radius, to avoid any surface effects in real space; and (\textit{ii}) $R$ should be maximum to still be considered as a local perturbation, in which no appreciable change in the Fermi level of the crystalline host can be observed. As our intention here is to analyze the implications of very short range chemical disorder in the torque-correlation theory for the Gilbert damping (in its nonlocal formulation), the $15$-atom cluster ($R=a_0$) is a simple and useful model. From now on, this model will be referred to as \textit{embedded cluster VCA} (or, in short, EC-VCA).

From the electronic structure of this spatially disordered alloy, we determine both the onsite (for $i=j$) and the non-local (for $i\neq j$) contributions of the Gilbert damping $\alpha_{ij}$ from the following expression,

\begin{equation}\label{eq:damping-dft}
\begin{split}
\alpha_{ij}^{\mu\nu}=\frac{g}{m_i \pi}\int_{-\infty}^{\infty}\eta(\epsilon)\textnormal{Tr}\left(\hat{T}_{i}^{\mu}\hat{A}_{ij}(\epsilon_{+})(\hat{T}_{j}^{\nu})^{\dagger}\hat{A}_{ji}(\epsilon_{+})\right)d\epsilon,
\end{split}
\end{equation}

\noindent where $\epsilon_{+}=\epsilon+i\delta$ for some small positive value $\delta$ (in energy units), $\textnormal{Tr}$ denotes the trace, the hat notation (e.g., $\hat{A}$) is used for operators (or tensors), and $\hat{A}_{ij}(\epsilon_{+})=\frac{1}{2i}(\hat{G}_{ij}(\epsilon_{+})-\hat{G}_{ji}^{\dagger}(\epsilon_{+}))$ is the $ij$ block of the spectral function at the energy $\epsilon$. The imaginary part $\delta$ describes the electron band broadening and, thus, is related to the electron lifetime associated with the relaxation from an excited state to the ground state due to intrinsic scattering events. This involves electro-magnon and electron-phonon scattering as well as electron correlation. 

Typically $\delta$ serves as the input, as seen in Ref.~\cite{PhysRevLett.99.027204} or is calculated as a self-energy by dedicated methods, such as the alloy analogy model used in \cite{PhysRevLett.107.066603}. However, in our approach, it is determined from the terminated continued fractions in the Haydock recursion method. It has been shown previously that this procedure results in a good agreement of the Gilbert damping for several materials to experimental measurements \cite{PhysRevB.108.014433}. The term $\hat{T}^{\mu}_i=\left[\hat{\sigma}^{\mu}_i,\hat{\mathcal{H}}_{so}\right]$ is the so-called torque operator \cite{PhysRevMaterials.2.013801} at site $i$,  $\mu,\nu\in\{x,y,z\}$ represent the Cartesian coordinates, and $\eta(\epsilon)=-\frac{\partial f(\epsilon)}{\partial\epsilon}$ is the derivative of the Fermi-Dirac distribution $f(\epsilon)$ with respect to the energy $\epsilon$. In all our simulations, the electron temperature is here taken to be zero, so the energy integral is only performed at the Fermi level. The total magnetic moment at a site $i$, which is equal to the sum of orbital moment $m^{\textnormal{orb}}_{i}$ and spin moment $m^{\textnormal{spin}}_{i}$,  is represented by $m_{i}$. The $g$-factor is given by $g=2\left(1+\frac{m^{\textnormal{orb}}}{m^{\textnormal{spin}}}\right)$ and $\hat{\sigma}^{\mu}_i$ is the $\mu$-component of the Pauli matrices vector (at site $i$).

  Altogether, Eq. \ref{eq:damping-dft} implies that the non-local Gilbert damping is generally non-symmetric with respect to the $ij$ pairs, as also pointed out by Ref.~\cite{PhysRevB.109.094417}: 

\begin{equation}
\label{eq:symmetry-alphaij}
\alpha_{ij}^{\mu\nu} = \left(\frac{m_j^{\textnormal{spin}}}{m_i^{\textnormal{spin}}}\right)\alpha_{ji}^{\nu\mu},
\end{equation}

\noindent which should hold because of the trace properties applied to Eq. \ref{eq:damping-dft}. Naturally, Eq. \ref{eq:damping-dft} results in a $3\times3$ damping tensor $\alpha_{ij}^{\mu\nu}$. From having the spin quantization axis along the $\mathbf{z}$ ($[001]$) direction, we then can define a scalar damping value $\alpha_{ij}$ as the average $\alpha_{ij}=\frac{1}{2}(\alpha_{ij}^{xx}+\alpha_{ij}^{yy})$ (for a discussion about the validity of this definition in a chemical disorder environment, please see Section \ref{sec:off-diagonal-elements}). 

From the non-local Gilbert damping, we define the effective damping of atomic site $i$, $\alpha_i$, as

\begin{equation}\label{eq:damping_sum}
\begin{split}
\alpha_i =\sum_j \alpha_{ij}.
\end{split}
\end{equation}

In practice, $\alpha_i$ is the cumulative non-local damping summation inside a given cutoff radius $r_c$ from the reference ($i$-th) site: $\alpha_i=\sum_{r_{ij}\leq r_c}\alpha_{ij}$. Since $\alpha_{ij}$ is typically long-range and scales asymptotically (i.e., for large $r_{ij}$) as $1/r_{ij}^2$ \cite{PhysRevMaterials.2.013801,PhysRevB.108.014433,PhysRevB.109.094417}, we consider the sum not only over the 15 atoms in the embedded impurity cluster, but also into the VCA medium in the EC-VCA model. For $r_c=6a_0$, $\alpha_i$ is approximately converged, involving $1836$ neighboring atoms in the damping calculations plus the onsite term. 

Experiments often measure damping as a single parameter from ferromagnetic resonance (FMR) techniques~\cite{ma2017metrology,Schoen2016}, in which the magnetic moments are excited in a coherent mode ($\mathbf{q}=\mathbf{0}$). As demonstrated in Appendix \ref{sec:damping_formula}, the measured value (\textit{effective} damping, $\alpha^{\textnormal{eff}}$) can be related to the site-resolved effective damping $\alpha_i$ (Eq. \ref{eq:damping_sum}) as

\begin{equation}
\label{eq:total-damping}
\alpha^{\textnormal{eff}}=\frac{1}{\sum_i m_i}\sum_i \alpha_im_i,
\end{equation}

\noindent where the index $i$ runs over all atoms in the unit cell. For the general case, however, the relation is unfortunately not so straightforward; a correspondent $\alpha^{\textnormal{eff}}$ should be possible to be extracted by comparing how the magnetization behaves dynamically in the non-local damping environment and in a local, single-parameter $\alpha$ candidate. We note that previous investigations have engaged with the aspect of an \textit{averaged} $\alpha$ from element- or site-specific dampings, albeit  
without using the moments $m_i$ as weights (e.g., Ref. \cite{PhysRevB.95.214417}). For an alloy, several configurations can be constructed for a given concentration of elements. In that case, the average of \textit{all} obtained $\alpha^{\textnormal{eff}}$ values (for that concentration and chosen $\delta$) is the best theoretical estimation for a measured intrinsic $\alpha$. This quantitative analysis for all concentrations of the FeCo system is not the primary aim here, although we investigate the full scenario for an Fe-rich (Fe$_{87}$Co$_{13}$) alloy in \ref{subsec:local-effects-detailed}. Instead, our emphasis is directed towards the impact of the chemical environment on the $\alpha_i$, aiming for the report of \textit{locally hidden} features of that quantity.

\subsection{Atomistic spin dynamics with an explicit embedded cluster}

The generalized LLG equation \cite{PhysRevB.109.094417,PhysRevMaterials.2.013801,PhysRevB.108.014433,PhysRevB.73.184427}, which incorporates the nonlocal damping, is solved by using an implicit midpoint solver as implemented in the UppASD code \cite{UppASD}, with a time step of $dt=10^{-17}$ s; for more details on the solver and the spin Hamiltonian used see Ref. \cite{PhysRevB.108.014433}. In each simulation, a $40\times40\times40$ spin lattice with perfect bcc crystal symmetry is modelled with periodic boundary conditions. To isolate the effects of distinct magnetic interactions and damping parameters, all calculations start from the same random (noncollinear) spin state. The spin-spin exchange interactions and the nonlocal damping parameters are considered up to a cutoff radius of $r_c=6a_0$, in compliance with the convergence criteria established in Section \ref{subsec:DFT}. As in Ref. \cite{PhysRevB.108.014433}, the fluctuation-dissipation relation in the presence of nonlocal damping is considered up to the approximation $\alpha_{ij}^{\mu\nu}\rightarrow\frac{1}{3}\textnormal{Tr}(\hat{\alpha}_{ij})\delta_{ij}\delta_{\mu\nu}$, which is strictly reasonable in the low-temperature limit ($T\rightarrow0$), i.e., for vanishing stochastic fields. To break the rotational invariance of the Heisenberg exchange interaction term, a small cubic magnetic anisotropy was also added to the spin Hamiltonian. This term is characterized by an experimental constant $K_1$ of the order of $\sim3$ $\mu$eV/atom and easy axis $[100]$, approximately correspondent to the Fe$_{87}$Co$_{13}$ composition \cite{Pfeifer1980}. As all calculations include such anisotropy value, it thus does not contribute to the final differences observed.

When an embedded cluster is incorporated (i.e., beyond just VCA-type interactions), its 15-atom center coincides with the spin lattice center. Both $J_{ij}$ and $\alpha_{ij}$ parameters correspondent to each explicit Fe or Co sites (up to a distance of $6a_0$) are included in the Hamiltonian, considering the symmetry rules $J_{ij}=J_{ji}$ and the one defined in Eq. \ref{eq:symmetry-alphaij}. Of course, when short-range disorder is explicitly considered, the inversion symmetry is also broken, and Dzyaloshinskii-Moriya interactions can emerge as a consequence. As Fe and Co are elements with relatively weak spin-orbit coupling (which is also a cause of the $\sim10^{-3}$ intrinsic damping values), we choose to disregard that term in the Hamiltonian and concentrate on the effects of the inhomogeneous nonlocal damping field.

\subsection{Selected clusters and notation}
\label{subsec:clusters}

In our study, the embedded nano-clusters within the EC-VCA model are tailored to match the composition of the surrounding VCA medium. It is important to realize, however, that in a real sample this boundary is less sharp, and local configurations of solely Fe or Co (alloy clustering) are possible, albeit being less likely; this scenario is deliberately avoided here. Due to the vast number of possible configurations, our study specifically investigates a total of 64 unique configurations across all the compositions. The study is particularly concentrated on investigating the influence of neighboring Co atoms' positions on the effective damping of the central atom within each cluster configuration. For simplicity, the $\alpha_i$ refers to the effective damping of the central (reference) atom $i$ in each cluster in the following discussion. 

\begin{figure*}[!ht]
        \includegraphics[width=2\columnwidth]{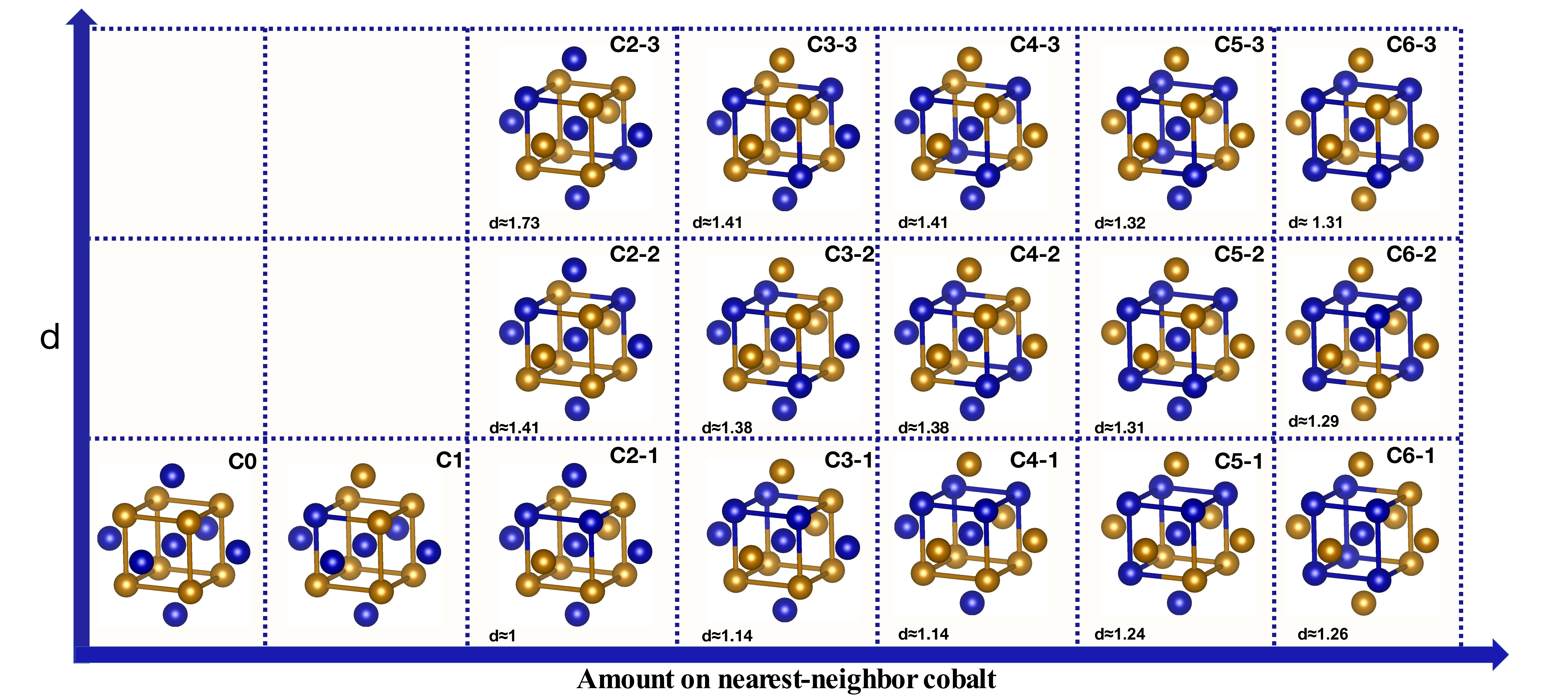} 
        \caption{
        Schematic representation of some possible embedded Fe$_{53}$Co$_{47}$ clusters in the VCA medium. Within this composition, there are 8 Fe atoms (yellow spheres) and 7 Co atoms (blue spheres) in total.
        The horizontal line organizes, from left to right, the embedded clusters with increasing amount on nearest-neighbor cobalt $n$ (see text). Within each column, the clusters have the same $n$ but an increasing average Co interatomic distance $d$ defined in Eq. \ref{eq:distance}(in units of $a_0$)  from the bottom to the top side.}
        \label{fig:cluster} 
\end{figure*}

As we deal with several configurations, an appropriate notation system is imperative. Figure \ref{fig:cluster} illustrates the set of investigated Co-centered clusters in the Fe$_{53}$Co$_{47}$ composition, in which the horizontal axis depicts, from left to right, the embedded clusters with increasing number of Co atoms at the first neighboring shell (1st NN) of the reference atom $i$. Whenever possible, the notation will be exemplified with that particular composition, even though it is transferable to other Fe$_{100-x}$Co$_{x}$ alloys. Therefore, we adopt the following definitions and conventions:

\begin{itemize}
    \item$n$ denotes the number of Co atoms positioned at the first neighboring shell of the reference (central) site. For Fe$_{53}$Co$_{47}$, $n=\{0,1,\ldots,6\}$. However, it may exceed these values for Co-enhanced compositions (e.g., Fe$_{40}$Co$_{60}$);
    \item $d$ symbolizes the average interatomic distance among the Co atoms in the first neighboring shell (normalized by lattice constant $a_0$), calculated as 
    \begin{equation}
\label{eq:distance}
d=\frac{1}{n(n-1)}\sum_{ij}\frac{|\boldsymbol{r}_{ij}|}{a_0},
\end{equation}
    where  $\boldsymbol{r}_{ij}$ is the distance vector between Co atoms at sites $i$ and $j$; 
    \item $m=\{1,2,3\}$ groups the clusters based on the value of their averages distances $d$. Thus, $m=1$ and $m=3$ correspond to the clusters with the smallest and largest $d$, respectively, whereas $m=2$ identifies clusters that fall in between;
    \item C$n$ labels the cluster with $n$ Co atoms in the first neighboring shell, for a given alloy composition;
    \item C$n$-$m$ categorizes the cluster C$n$, identifying it with the proper Co average distance group $m$ (not used for the cases C0, C1, C7, and C8 by construction).
\end{itemize}
 
It should be noted that the possible number of average distances groups $m$ is larger than the 3 shown here. We have confined the number of clusters due to considerations of computational efficiency. Despite this constraint, the selected cases can adequately demonstrate the underlying physics. The investigated clusters can have the same 1st NN Co arrangement but a different 2nd NN environment in different compositions, attributed to the varying Co amounts. The total number of Co atoms and the maximum/minimum $n$ of each composition are shown in Table \ref{tab:cluster} in Appendix \ref{sec:Cluster_configuration}. 

\section{RESULTS}
\label{sec:results}

\subsection{Investigation of off-diagonal elements}
\label{sec:off-diagonal-elements}

Before proceeding to the analysis of local chemical disorder effects on the magnetic damping, a natural question that arises -- linked to the tensorial character of $\alpha$ -- is about the practical contribution coming from the off-diagonal elements in a heterogeneus environment. In pure bcc or fcc $3d$ ferromagnets, when the spin quantization axis (SQA) aligns with a fourfold or threefold symmetry axis (e.g., in $\mathbf{z}$ direction), only diagonal elements are expected to be non-zero in the Gilbert damping tensor, as discussed in Refs. \cite{eriksson2017atomistic,PhysRevB.72.064450,PhysRevMaterials.2.013801} and corroborated by \cite{PhysRevB.108.014433}. This situation becomes ideal to extract a scalar $\alpha_{ij}$, which can be defined as $\alpha_{ij}=\frac{1}{2}\left(\alpha_{ij}^{xx}+\alpha_{ij}^{yy}\right)$, and to directly compare to what is typically measured in experiments \cite{PhysRevB.108.014433}. However, when considering an alloy with explicit chemical disorder, the intrinsic symmetry is broken, rendering non-zero off-diagonal elements plausible, even with SQA$\parallel\mathbf{z}$.

To address this problem, we considered the effective off-diagonal terms $\alpha^{\textnormal{eff}}_{xy}$ and $\alpha^{\textnormal{eff}}_{yx}$ (see Eq. \ref{eq:effective-tensor-elements}) from a bcc special quasirandom structure containing 16 atoms (SQS-16) \cite{Zunger1990}, obtained for a particularly simple FeCo concentration (Fe$_{50}$Co$_{50}$). The SQS-16 cell was created using the MCSQS algorithm within the framework of the Alloy Theoretic Automated Toolkit (ATAT) \cite{vandeWalle2013}; more details on the method inputs to create the cell can be found in Ref. \cite{PhysRevB.108.014433}.

Clearly, the off-diagonal terms do not vanish (see Fig. \ref{fig:off-diag-sqs} in the Appendix \ref{sec:off-diagonal-convergence}), but they represent a very small fraction of the diagonal terms ($\sim4-5\%$). Also, the asymmetry is reduced between the diagonal terms ($\sim2\%$), which indicates that at large scales the chemical disorder plays a smaller role on the format of the $\hat{\alpha}^{\textnormal{eff}}$ tensor than the symmetry of the bcc structure in connection with the SQA itself. In other words, it is a decent approximation to consider $\alpha^{\textnormal{eff}}\rightarrow\frac{1}{2}(\alpha_{xx}^{\textnormal{eff}}+\alpha_{yy}^{\textnormal{eff}})\sim\alpha_{xx}^{\textnormal{eff}}$. Therefore, although locally off-diagonal elements maybe not negligible, we can consider in our investigation only the average $\alpha_{ij}=\frac{1}{2}\left(\alpha_{ij}^{xx}+\alpha_{ij}^{yy}\right)$, that will effectively contribute to $\alpha^{\textnormal{eff}}$ in the bulk scale.

\subsection{Local effects: the Fe$_{87}$Co$_{13}$ case}
\label{subsec:local-effects-detailed}

Contrary to the Heisenberg exchange, which shows an isotropic nature in the collinear state (i.e., dependent on the atomic pair distance, and without the SQA as a symmetry reducer), the Gilbert damping is anisotropic with respect to the SQA -- even within the same neighboring shell. Formally, while the former is explained by a simultaneous conservation of the trace of the block Green's functions $\hat{G}_{ij}$ (and, consequently, also $\hat{A}_{ij}$) and $\ell$-dependent diagonal exchange-splitting matrices (see, e.g., \cite{PhysRevB.62.5293,PhysRevB.91.125133}), the later is explained by the non-diagonal character of the torque operators $\hat{T}_i^{\mu}$ (Eq. \ref{eq:damping-dft}). In other words, not only the distance and chemical classification of the $ij$-pair is important for the composition of $\alpha_{ij}$ values, but also their position with respect to the SQA (throughout the text we consider $\textnormal{SQA}\parallel\mathbf{z}$ -- unless otherwise noted). Physically, this can be understood by two connected facts: (\textit{i}) for small rotations of the spin about the SQA ($\mathbf{z}$), if it is a high symmetry axis,  the change in the magnetization will stay confined in the other two independent reference orientations ($\mathbf{x}$-$\mathbf{y}$ plane in this case, so $\mu\in\{x,y\}$), and (\textit{ii})  the spin-orbit torques will recruit different contributions from the orbitals in the inter-site electronic spectral functions (i.e., through $\hat{A}_{ij}$) to participate in the dissipation process, obeying their symmetries in real-space. This anisotropic behavior of $\alpha_{ij}$ has been observed and reported in Refs. \cite{PhysRevB.103.L220405,PhysRevB.108.014433,PhysRevMaterials.2.013801}, where sometimes more than a single value of non-local damping is found for the same inter-site distance $r_{ij}$, even for single-element materials. As an example, we can analyze the second neighboring shell of (pure) bcc Fe, which presents two distinct $\alpha_{ij}$ values \cite{PhysRevB.108.014433}. In this case, the orbital symmetry of the sites localized at positions $(\pm a_0,0,0)$ and $(0,\pm a_0,0)$ ensure a swap between $\alpha_{ij}^{xx}$ and $\alpha_{ij}^{yy}$ in the respective $\hat{\alpha}_{ij}$ tensors, resulting in the same $\alpha_{ij}$ parameters according to the definition in Section \ref{sec:methods}. However, the sites localized at positions $(0,0,\pm a_0)$ are characterized by inter-site propagators $\hat{G}_{ij}$ with completely different (diagonal and non-diagonal) orbital contributions, resulting also in distinct $\hat{\alpha}_{ij}$ tensors when compared to the previous 4 sites. Unfortunately, the non-diagonal character of the $\hat{T}_i^{\mu}$ operators prevents us from discretizing the diagonal $\alpha_{ij}^{\mu\nu=\mu}$ elements in orbital contributions in a simple and analogous way as for the Heisenberg exchange interactions ($J_{ij}$'s, see Refs. \cite{PhysRevB.91.224405,PhysRevLett.116.217202}).

Given the complexity associated with the set of degrees of freedom that potentially influence the real-space $\alpha_{ij}$'s (chemical classification, distance, and position around a SQA), the best possible scenario to investigate local effects is to consider all configurations for the particular alloy concentration of interest. In this context, as a bonus, not only local effects can be studied, but also explicit alloying disorder (short-range disorder) can be indirectly considered as one source of electronic finite lifetime in these materials \cite{PhysRevB.87.014430} -- generally treated by effective medium approaches, such as the simple VCA \cite{PhysRevB.108.014433} or the more accurate CPA \cite{PhysRevB.87.014430,PhysRevB.98.174412,PhysRevLett.105.236601,PhysRevB.92.214407}.  In a 15-atom cluster, with $p$ number of Co atoms inside the cluster, this means a total number of configurations equal to $C^{15}_{p}=\binom{15}{p}=\frac{15!}{p!(15-p)!}$ for a given concentration Fe$_{{100-\frac{20p}{3}}}$Co$_{\frac{20p}{3}}$.

As this result scales quickly to thousands of possible configurations, we can choose an Fe-rich alloy to be our proof-of-concept, i.e., we choose $p=2$ (or Fe$_{87}$Co$_{13}$). In this case, the 105 total configurations can be reduced to 16 nonequivalent situations, following the symmetry of the Gilbert damping parameters in the bcc structure. However, before we analyze the average on this alloy, we discuss in detail a single case among the 16 nonequivalent situations.

\begin{figure}[!ht] 
\centering
        \includegraphics[width=0.6\columnwidth]{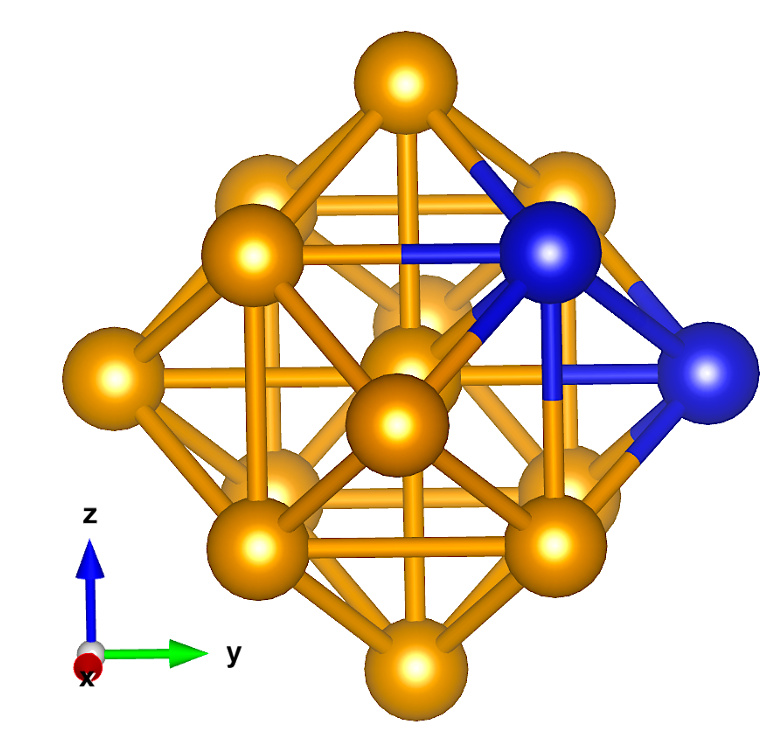}
        \caption{(Color online) Schematic representation of a 15-atom embeddedcluster with alloy concentration Fe$_{87}$Co$_{13}$. Within the EC-VCA model, this figure depicts only the explicit part, which in the calculations is surrounded by the corresponding VCA medium. The color scheme of the spheres is consistent with Fig. \ref{fig:ec-vca-representation}. Bonds are drawn to guide the eye. Image produced with the VESTA software \cite{Momma2011}.}
        \label{fig:cluster-example} 
\end{figure}

As an example, consider the cluster displayed in Figure \ref{fig:cluster-example}. In principle, due to the cubic environment, each of the $\{\mathbf{x},\mathbf{y},\mathbf{z}\}$ axes present a 4-fold ($C_4$) symmetry with respect to the $\alpha_{ij}$ values, representing a total of 24 non-equivalent configurations (from the initial 105). However, in view of the reasons discussed above, when a Co atom occupies any site within the SQA ($\mathbf{z}$-axis), the $\alpha_{\textnormal{Fe-Co}}$ value is distinct, reducing the multiplicity of this example to 16. In that case, the variation of the nearest-neighbor Co position w.r.t. the central Fe atom do not change the $\alpha_{ij}^{\mu\nu}$ components, but the second nearest-neighbors (NNN), even when occupied by Co atoms, follow the same symmetry rules as the pure bcc Fe (swap $\alpha_{ij}^{xx}\leftrightarrow\alpha_{ij}^{yy}$ for NNN in the $xy$-plane, different components for NNN in $\mathbf{z}$). 

We can quantify the difference for this example.  Taking into account the multiplicity, the variation in $\alpha_{i}$  (Eq. \ref{eq:damping_sum}, $r_c=6a_0$) of having Co in the $\mathbf{z}$-axis or in the $xy$-plane amounts to $\sim1\%$ -- similarly to other quasi-equivalent cases among the 16 configurations. That is not a purely non-local effect, as the onsite contribution also varies with about the same intensity. Although small for Fe-Co alloys, given their relatively weak spin-orbit strength, we anticipate that the variation in $\alpha_{i}$ is larger for environments with higher spin-orbit coupling, $\xi$, in the limit where the adiabatic and mean-field approximations (assumed for torque-correlation formalism used here) are still reasonable \cite{PhysRevB.92.014419}.

\begin{figure*}[!ht] 
\centering
        \includegraphics[width=\textwidth]{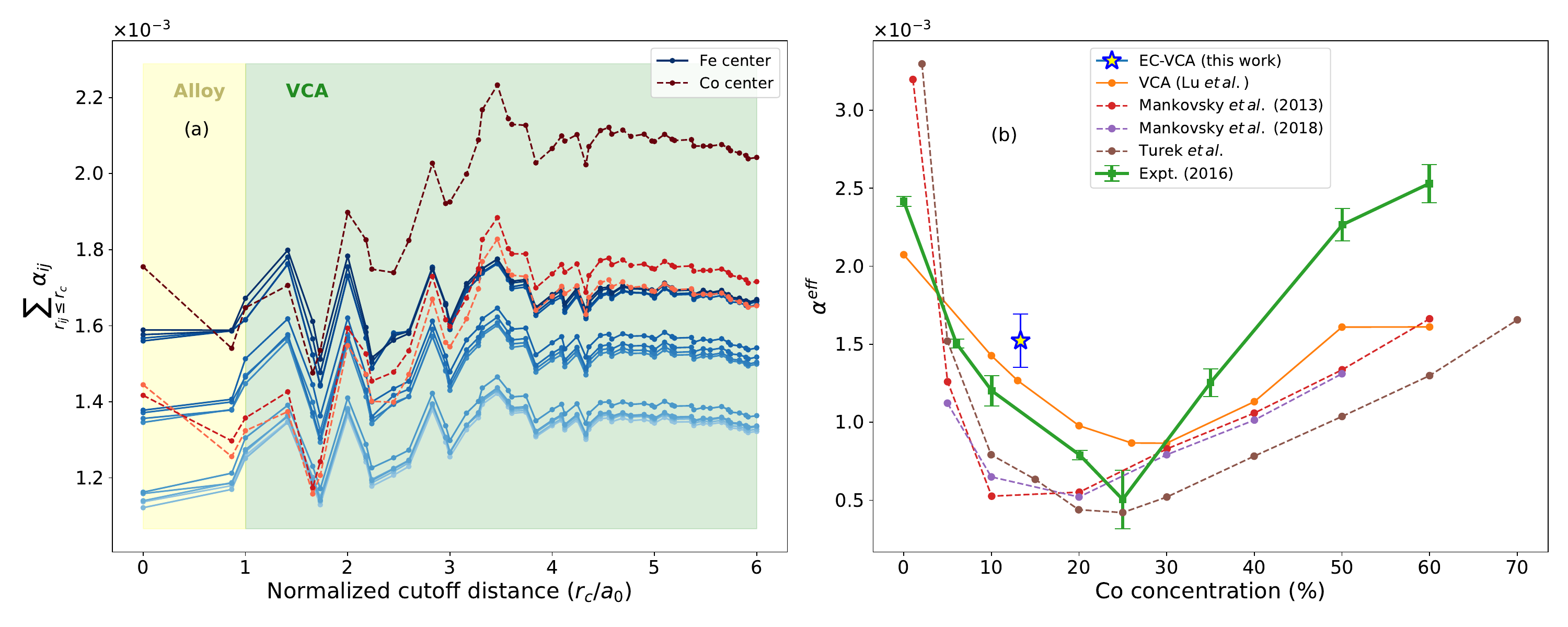}
        \caption{(a)  $\sum_{r_{ij}\leq r_c}\alpha_{ij}$ up to a given cutoff radius ($r_c$) of all nonequivalent configurations of Fe$_{87}$Co$_{13}$ in the EC-VCA model. The different curves shown in colors with different shades of blue represent different Fe-centered clusters, while the curves shown with different shades of red show different Co-centered ones. (b) Comparison of the effective damping obtained from all Fe$_{87}$Co$_{13}$ configurations in the EC-VCA model (blue point) with theoretical values obtained by the use of effective medium approaches (VCA \cite{PhysRevB.108.014433}, and CPA \cite{PhysRevB.92.214407,PhysRevB.87.014430,PhysRevB.98.104406}) and with the experimental values of intrinsic damping \cite{Schoen2016}. In (a), the yellow and green areas represent the explicit alloy and VCA medium regions of the EC-VCA model, respectively.
        } 
        \label{fig:damping-fe87co13} 
\end{figure*}

It is instructive to analyze the damping that can be calculated from all possible nonequivalent configurations of Fe$_{87}$Co$_{13}$. The results of both individual cumulative damping summations (Eq. \ref{eq:damping_sum}), and comparison between the effective damping of Fe$_{87}$Co$_{13}$ and previous literature results are shown in Figure \ref{fig:damping-fe87co13}. We first notice from Fig. \ref{fig:damping-fe87co13}(a) that generally, the values of $\alpha_i$ for Co-centered clusters in the explicit alloy region are higher (scaling with $\frac{1}{m_i}$) than the Fe-centered ones, and also subject to more oscillations, especially throughout the VCA medium. More importantly, even within each set of Fe- or Co-centered configurations, we note the presence of distinct curves, which is a direct consequence of the different spatial distributions of Fe/Co inside the embedded cluster region -- or, in other words, a direct influence of the local environment. Although the onsite contribution ($\alpha_{ii}$) drives the majority of the effect, essentially due to changes in the local density of states around $E_F$ because of the charge transfer between species, the non-local contribution cannot be disregarded. This is partially in line with the importance of site-resolved non-local damping suggested in Ref. \cite{PhysRevB.109.094417}, although the authors indicate it in the context of reduced dimensionality -- and here we argue its relevance in the presence of short-range chemical disorder. Explicitly, the largest variation of the $\alpha_{ii}$ values among the Fe- and Co-centered clusters are $\sim4.7\times10^{-4}$ and $\sim3.4\times10^{-4}$, respectively, while the (pure) nonlocal contributions vary $\sim0.9\times10^{-4}$ ($\sim1.3\times10^{-4}$) and $\sim0.6\times10^{-4}$ ($\sim0.9\times10^{-4}$) for $r_c=a_0$ ($r_c=6a_0$), where $r_c=a_0$ represents the limit of the embedded cluster region.

With the definition given by Eq. \ref{eq:total-damping}, the effective Gilbert damping can be obtained. The comparison of $\alpha^{\textnormal{eff}}$ for Fe$_{87}$Co$_{13}$ with other literature results is shown in Fig. \ref{fig:damping-fe87co13}(b). The error bar, here identified as the precision of $\alpha^{\textnormal{eff}}$, is calculated by the standard deviation and evaluated as $\sim11\%$. This percentage will be used in a later stage as the approximate error of other Co concentrations averages. The final result of the EC-VCA model, $\alpha^{\textnormal{eff}}=(1.52\pm0.17)\times10^{-3}$, can then be compared to the pure VCA calculation, $\alpha^{\textnormal{eff}}=1.27\times10^{-3}$. Despite the consideration of almost the same chosen broadening parameter, $\delta$, the effective damping for the EC-VCA model is enhanced in comparison with the pure VCA value. This is partially related to the fact that the configuration average indirectly introduces the chemical disorder as a cause for change in $\delta$ (electronic lifetime), and the randomly arranged atoms as centers for electronic scattering. Another obvious reason associated with this difference, that cannot be neglected, is the relative small size of the embedded cluster. However, as Fig. \ref{fig:damping-fe87co13}(b) shows, both the VCA and EC-VCA calculatons capture most experimental results fairly well.

\subsubsection{Spin dynamics: remagnetization}
\label{sec:explicit-asd}

The calculation of explicit $\alpha_{ij}$ parameters in the EC-VCA model allows for the inspection of the local environment effects on dynamical processes, in a fully atomistic picture. 
Although the impact of a 15-atom explicit region on the overall energy dissipation rate is expected to be small, the spin dynamics within this region can significantly differ from that obtained considering an array consisting solely of VCA sites. In this sense, the remagnetization stands as a suitable process to investigate; its relevance becomes evident in the context of, e.g., ultrafast pump-probe experiments and magnetization switching \cite{Radu2011,PhysRevLett.121.077204,Pankratova2024}, even though care should be taken in the direct comparison to those experimental results as variations in the spin moment length are not accounted for in the generalized LLG equation \cite{PhysRevB.109.094417,PhysRevMaterials.2.013801,PhysRevB.108.014433,PhysRevB.73.184427}.

With this aim, we assume minimal influence from off-diagonal elements (see Section \ref{sec:off-diagonal-elements}) and employ the method outlined in Ref. \cite{PhysRevB.108.014433}. As a test case, we examine the remagnetization dynamics with an embedded cluster as shown in Fig. \ref{fig:cluster-example}, starting from a random noncollinear state (same for all simulations). This dynamics with an explicit cluster can then be compared to the Fe$_{87}$Co$_{13}$ fully VCA model to account for the impact of the locally distinct nonlocal Gilbert dampings. In this context, the $J_{ij}$ set also differs. Thus, a third model in which all exchange interactions are artificially replaced by the corresponding VCA values is considered, to isolate the effect of $\alpha_{ij}$'s. All sets of computed $J_{ij}$ parameters, correspondent to both the pure VCA calculation and the EC-VCA model, are depicted in Appendix \ref{sec:jijs-sets-asd}.

Figure \ref{fig:explicit-asd}(a) shows the obtained average magnetic moments, considering both nonlocal and site-resolved effective damping descriptions, for all three models (pure VCA, EC-VCA, and EC-VCA with the VCA $J_{ij}$ set). The first point to be noted is the global decrease of the energy dissipation rate by the nonlocal damping formulation of the equation of motion ($\alpha_{ij}$, full lines) compared to the effective damping ($\alpha_i$, dashed lines). Moreover, all three models show distinct remagnetization times, as highlighted by the \textit{Insets}. For example, among the three models, we obtain maximum differences  of $\sim600$ fs and $\sim200$ fs to reach $\sim77\%$ of the fully saturated magnetization in the $\alpha_i$ and $\alpha_{ij}$ simulations, respectively; as these differences occur between the pure VCA and the EC-VCA with the same exchanged $J_{ij}$ set, they can be ascribed solely to the $\alpha_{ij}$'s. Interestingly, here the contrast between the nonlocal and effective descriptions is also evident: the dissipation rate of the pure VCA model is the highest in the former and the lowest in the latter.

Altogether, those results clearly demonstrate how the short-range disorder influence on the local dynamics can modify the overall relaxation rate. However, here the impact is reduced due to two main factors: (\textit{i}) the previously mentioned small replaced region in the VCA matrix (15-atom cluster in a $40\times40\times40$  lattice); and (\textit{ii}) the choice of a cluster whose effective damping (central atom $\alpha_i=1.32\times10^{-3}$) does not considerably deviate from the VCA value ($\alpha_i=1.27\times10^{-3}$). This suggests promising prospects for enhanced effects in larger embedded regions or in scenarios where damping varies significantly (up to an order of magnitude), as explored in the next section. It also corroborates with the assertion from Ref. \cite{PhysRevB.109.094417} that considering inhomogeneous damping can improve the accuracy of spin dynamics simulations.

To better illustrate the influence of chemical disorder on the local dynamics, in Fig. \ref{fig:explicit-asd}(b) we show the trajectories of both the central spin, which coincides with the center of the embedded cluster, and a peripheral spin located more than $6a_0$ from the center, encompassing only VCA-type interactions. Concerning the former, it is possible to see that the trajectories largely deviate in all three models from the beginning. In turn, for the latter, the dynamics only start to differ after a time interval of $\Delta t\sim 0.13$ ps, by the perturbation of spin waves coming from the cluster region.

\begin{figure*}[!ht] 
\centering
        \includegraphics[width=0.9\textwidth]{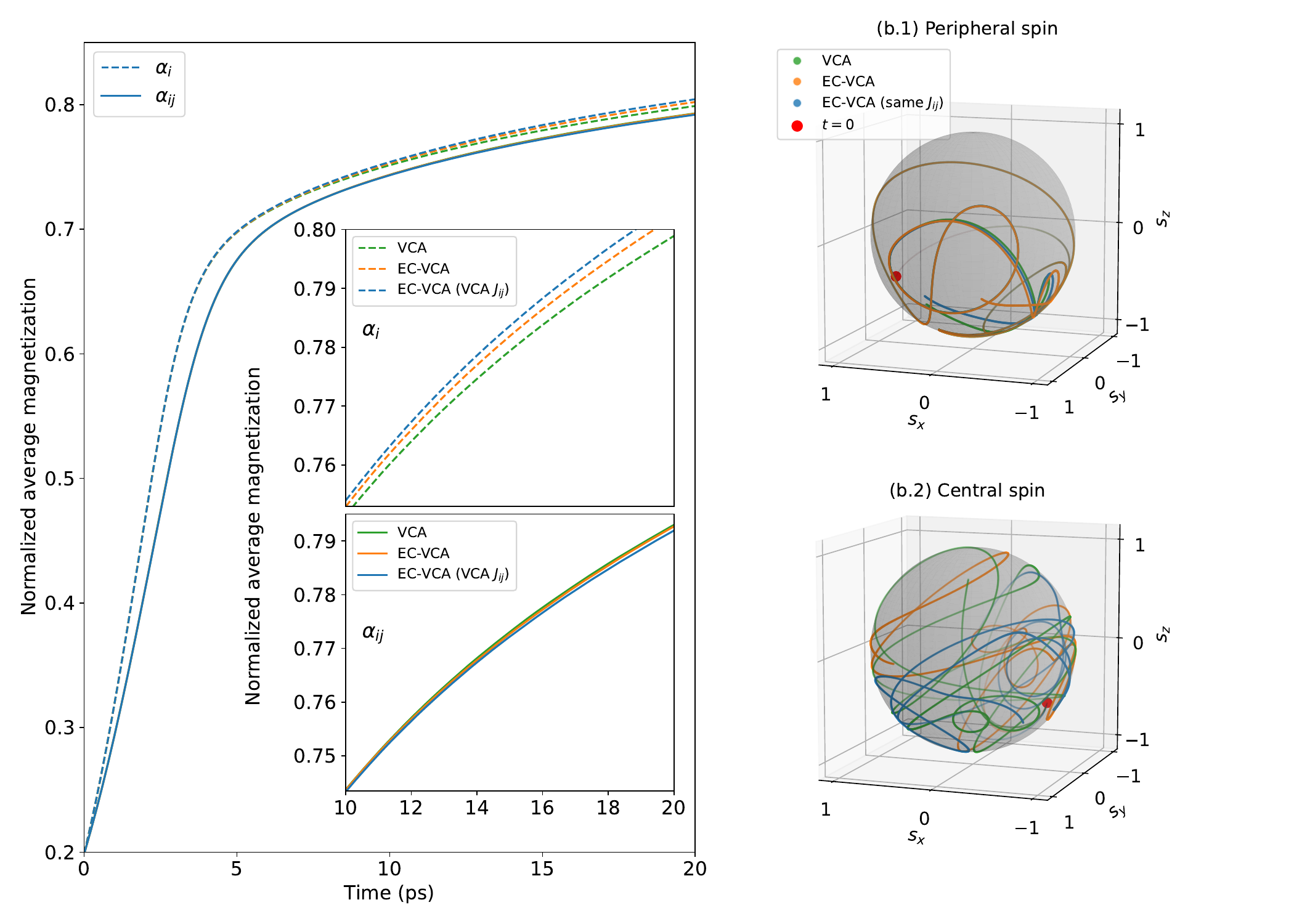}
        \caption{(a) Remagnetization process for Fe$_{87}$Co$_{13}$ simulated with atomistic spin dynamics, considering both the fully nonlocal Gilbert damping ($\alpha_{ij}$, solid lines) and the site-resolved effective damping ($\alpha_i$, Eq. \ref{eq:damping_sum}, dashed lines). Green, orange and blue lines symbolize the simulations performed with purely VCA, EC-VCA, and EC-VCA parameters considering the VCA $J_{ij}$ set, respectively. \textit{Insets:} a zoom-in of the normalized magnetization curves. (b) Spin trajectories lying on the unit sphere at $\Delta t\leq 0.13$ ps for (b.1) a peripheral and (b.2) the central spin, which coincides with the center of the embedded cluster. The red dot in (b.1) and (b.2) marks the initial position of the spin (at $t=0$), equivalent for the three models considered. 
        }
        \label{fig:explicit-asd} 
\end{figure*}

\subsection{Short-range disorder impact on damping: a view throughout distinct alloy concentrations}
\label{subsec:damping variation}

Having previously demonstrated and quantified the impact of short-range disorder on the effective Gilbert damping and spin dynamics by fully analyzing a particular FeCo concentration (Fe$_{87}$Co$_{13}$), in this section we extend our investigation to include all concentrations of Co (up to $60\%$), with a focus on the damping behaviour with respect to average Co interatomic distances, $d$.

The interatomic separation within clusters has a considerable influence on magnetic properties, such as magnetic exchange and magnetic moment, as already identified in previous studies \cite{PhysRevB.68.035207,doi:10.1021/acs.jpca.8b00540,peters2016magnetism}, but its impact on the effective damping is still poorly understood. 
To further explore the impact of Co interatomic distances on $\alpha_i$, we keep a constant number of Co atoms at the 1st NN sites($n$) along with a consistent 2nd NN atomic environment across each cluster group. We then vary the arrangement of the Co atoms at the nearest neighborhood around the central (or reference) atom, thereby altering the average distance $d$ defined in Section \ref{subsec:clusters}. It is important to note that multiple cluster configurations, not necessarily equivalent in terms of damping values, can share the same $n$ and $d$ values. For instance, fixing the 2nd NN environment for C2-1 results in 12 possible clusters per composition. However, a single cluster is randomly selected from these as a representative to analyze its damping. Consequently, our analysis is limited to these selected clusters.
It should be further noted that as the cluster groups C0, C1, C7, and C8 are comprised of a single cluster each (i.e., no variation in $d$), our investigation predominantly employs the C2 to C6 cluster groups as a platform.

To analyze the effective damping variation among clusters, we introduce two quantities: $\Delta\alpha^{\textnormal{eff}}$, defined, for each concentration $x$, by the clusters within the same C$n$ group with the smallest ($m=1$) and largest ($m=3$) $d$, set as 
\begin{equation}\label{eq:damping_variation}
\Delta\alpha^{\textnormal{eff}}=\frac{(\alpha_i^{m=3}-\alpha_i^{m=1})}{\alpha_i^{m=1}} \times 100\%\:\:, 
\end{equation}
and $\Delta\alpha^{\textnormal{max}}$, which represents the highest variation among all clusters investigated within each alloy concentration (not necessarily with same $n$ or $m$ indexes).

\begin{figure}[!ht]
\centering
        \includegraphics[width=0.9\columnwidth]{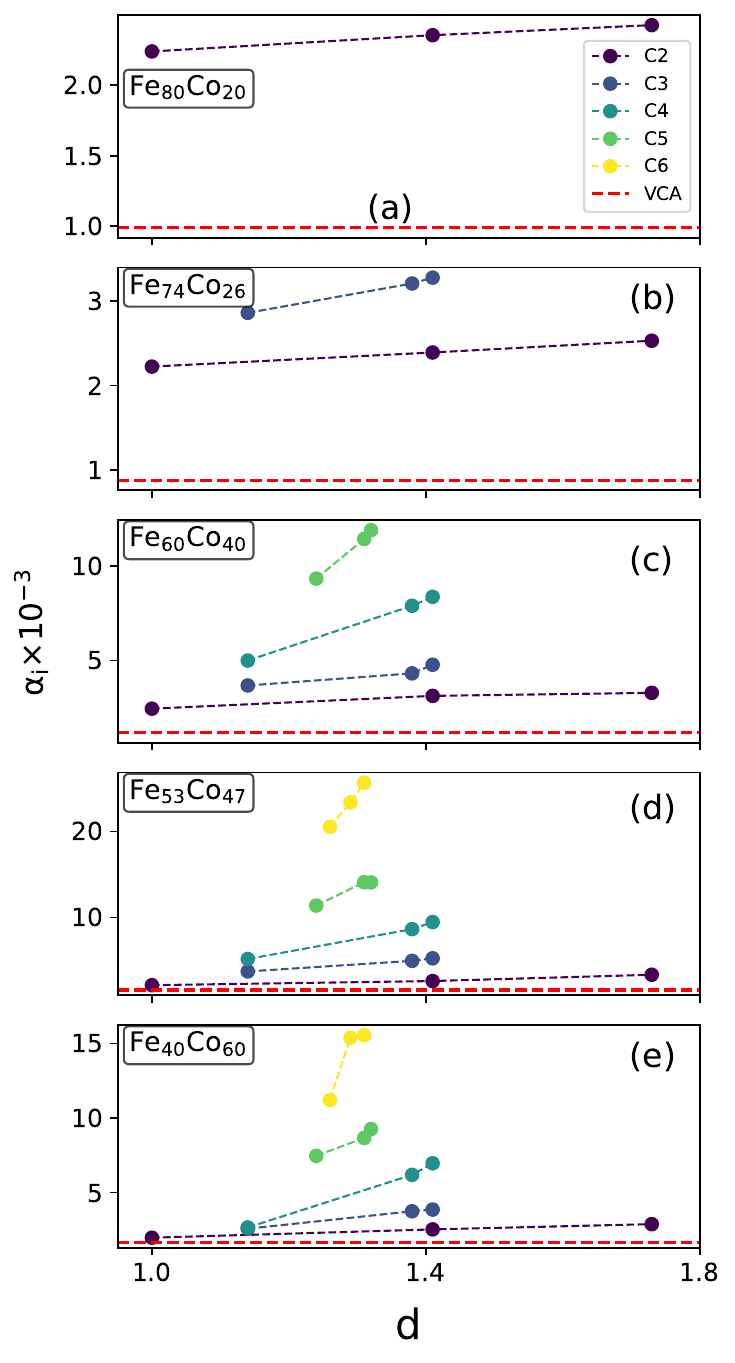}
        \caption{
        Site-resolved damping parameter, $\alpha_i$ as a function of the normalized average Co interatomic distance $d$ as defined in Eq. \ref{eq:distance}, in units of $a_0$.     The filled circles in (a) to (e) show the effective damping of the clusters within C2-C6 groups for different FeCo alloy compositions. As a reference, the red dashed line shows the damping computed considering purely the VCA model.} 
        \label{fig3:alpha_distance} 
\end{figure}

\begin{table}[h]
\caption{Effective Gilbert damping variation $\Delta\alpha^{\text{eff}}$ as a function of \(n\) and composition \(x\) of Fe$_{100-x}$Co$_{x}$. $\Delta\alpha^{\text{max}}(\%)$, $\Delta n({E_f})^{\text{max}}(\%)$ represent the maximum variation of damping as well as density of states at the Fermi level between clusters within each Co composition, respectively.}
    \centering
    \begin{tabular}{cccccc}
        \toprule
        \toprule
    $n$ & & &$x$ $(\%)$ & & \\
         \toprule
           &   20 & 26 & 40 & 47 & 60 \\
        \midrule
        2 & 8 & 14 & 34 & 58 & 46 \\
        3 & -- & 15 & 30 & 41 & 50 \\
        4 & -- & -- & 68 & 83 & 161 \\
        5 & -- & -- & 27 & 24 & 24 \\
        6 & -- & -- & -- & 25 & 39 \\
        \midrule
        $\Delta\alpha^{\text{max}}$ & 8 & 47 & 387 & 1122 & 685 \\
        $\Delta n({E_f})^{\text{max}}$ & 3 & 22 & 115 & 195 & 150 \\
        \bottomrule
        \bottomrule
    \end{tabular}
    \label{tab:effdampvar}
\end{table}

Figure \ref{fig3:alpha_distance} shows $\alpha_i$ as a function of $d$, in different cluster groups C$n$ and alloy concentrations $x$. From those results, two observations can be made: first, $\Delta\alpha^{\textnormal{eff}}$ is in all cases positive and can reach values up to $161\%$ in C4 cluster group of Fe$_{40}$Co$_{60}$. Consequently, damping tends to increase with $d$ (see Table \ref{tab:effdampvar}).  

Second, the damping parameter depends on $n$, indicating a shift towards higher damping magnitudes as more Co atoms occupy 1st NN sites. Comparing the lowest and highest damping value among $n$ and $m$ for a fixed composition $x$, we observe a variation of up to $1122\%$ (or one order of magnitude) for $x=47$ between the configurations C2-1 and C6-3 (see Table \ref{tab:effdampvar}, bottom row, and Fig. \ref{fig:cluster} for an illustration of the referred clusters). This corroborates with the significance of the 1st NN Co atoms on the effective Gilbert damping in the central (or reference) site of the embedded cluster.

\begin{figure}[!ht]
\centering
        \includegraphics[width=1\columnwidth]{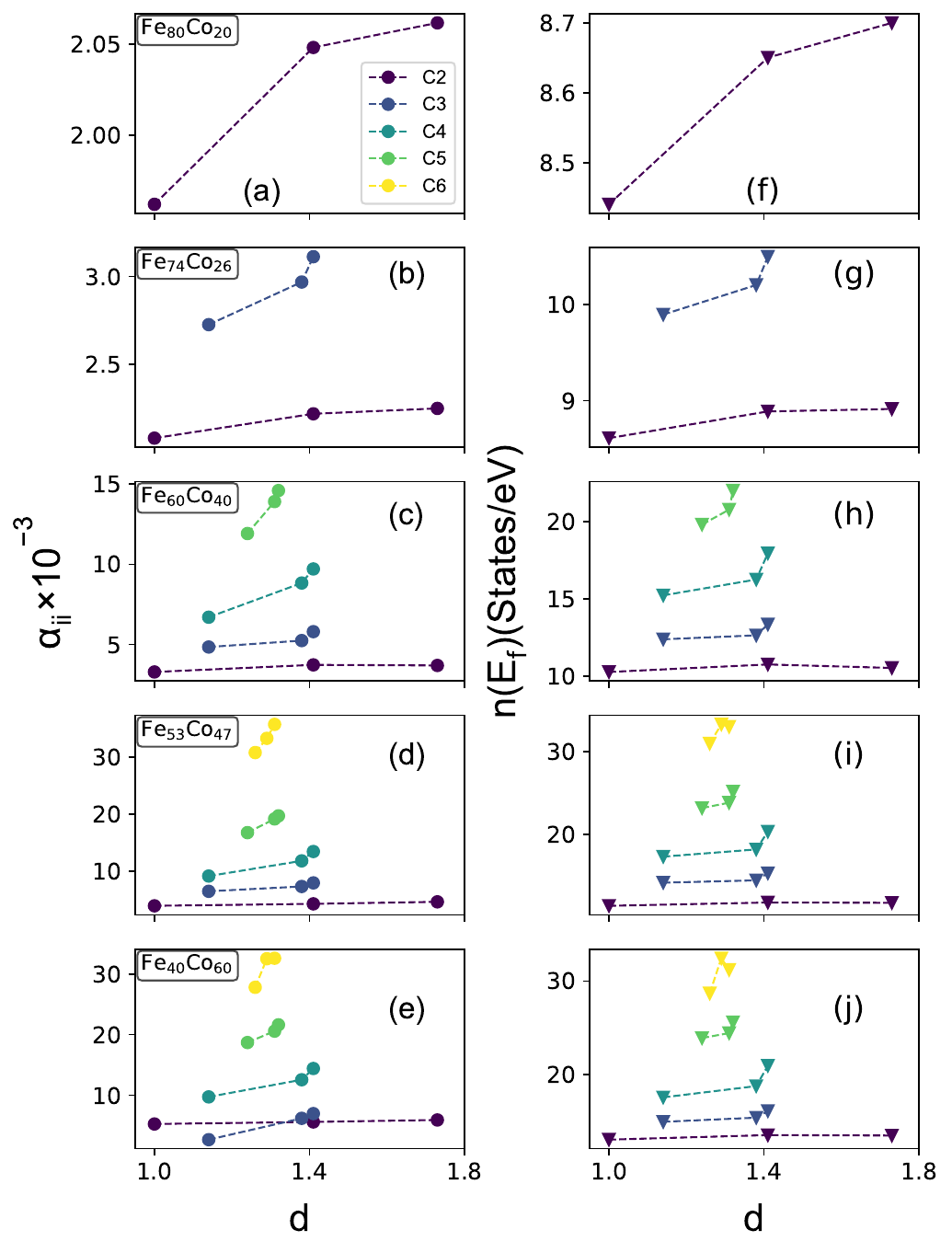}
        \caption{Onsite damping $\alpha_{ii}$ (left panel) and density of states at the Fermi level $n(E_f)$ (right panel) as a function of the normalized average Co interatomic distance $d$ as defined in Eq. \ref{eq:distance}, in units of $a_0$, for different FeCo alloy compositions. The colors follow the scheme of Fig. \ref{fig3:alpha_distance}.}
        \label{fig:onsite_dos} 
\end{figure}

To delve deeper into damping's dependence on $d$, and to understand the relevant enhancement of damping in terms of the number of 1st NN Co atoms, we analyze the on-site $\alpha^{\textnormal{onsite}}$ (see Fig. \ref{fig:onsite_dos}, left column) and the nonlocal $\alpha^{\textnormal{nonlocal}}$ contributions. We obtain these two parameters from decomposing Eq. \ref{eq:damping_sum} into  $\alpha^{\textnormal{onsite}}=\alpha_{ii}$ and $\alpha^{\textnormal{nonlocal}}=\sum_{i\neq j}\alpha_{ij}$. In agreement with previous investigations \cite{PhysRevB.103.L220405, PhysRevMaterials.2.013801,PhysRevB.108.014433}, $\alpha^{\textnormal{onsite}}$ is positive and typically larger than $\alpha^{\textnormal{nonlocal}}$ for all clusters; for $3d$ materials such as FeCo, each $\alpha_{ij}$ is one order of magnitude lower than $\alpha_{ii}$, or more. Analogously to the effective damping, $\alpha^{\textnormal{onsite}}$ tends to increase with $d$ as well as $n$ and with similar variations. Surprisingly, $\alpha^{\textnormal{nonlocal}}$ changes sign and starts to be negative at $x>40\%$ for all $n$ and $m$. Different to the damping from pure VCA, where $\alpha^{\textnormal{nonlocal}}$ turns negative only at $x=60\%$, the nearest neighbor $\alpha_{ij}$'s drive the most significant contribution to the negative $\alpha^{\textnormal{nonlocal}}$ at $x>40\%$ (data not shown). Furthermore, $\alpha^{\textnormal{onsite}}$ seems to be more sensitive (by a factor of $\sim10$) in variations of $n$, $d$, and $x$ compared to $\alpha^{\textnormal{nonlocal}}$ and, thus, serves as the dominant component in the effective damping trends. The only exception to this general observation is the C2 cluster group, where onsite and nonlocal dampings are equally sensitive in variations of $d$ and for all Co concentrations $x$.

To further identify the electronic structure origin of the results in the effective, onsite, and nonlocal dampings, we link the first principles determined dampings to the approximate (or simplified) Kambersky’s formula for the relaxation of spin-dependent electronic populations  \cite{Kambersky1970},

\begin{equation}\label{eq:kambersky}
\begin{split}
\alpha = \frac{\gamma}{4 M_s} n({E_f})\xi^2(\delta g)^2\tau ,
\end{split}
\end{equation}
where $\gamma$ is the gyromagnetic ratio, $ M_s$ denotes the saturation magnetization, $n({E_f})$ represents the total density of states at the Fermi level, $\xi$ is the strength of spin-orbit coupling (SOC), $\delta g$ is the deviation of the $g$-factor from the free-electron value, 
and $\tau$ is the electron lifetime related to intrinsic scattering events with other degrees of freedom. 

The variation of the damping with respect to possible changes in $\xi$ is  negligible across all cluster groups, with the largest difference of less than $1\%$ found for the $3d$ states contribution to the SOC constant. 
Regarding $\tau$, its relation to the terminator and recursion levels of the Haydock method shows minor differences among clusters \cite{PhysRevB.103.L220405}, although the precise calculation of $\tau$ is beyond the scope of this study. In turn, the damping of different ferromagnetic materials have been associated with $(\delta g)^2$ before \cite{Oogane2006}, but the $\frac{m^{\textnormal{orb}}}{m^{\textnormal{spin}}}$ relation, within Kittel's first-order perturbation picture \cite{Shaw2021}, only changes subtly in each concentration $x$ (and specially within each C$n$ group). Explicitly, in all considered clusters, the central Co exhibits a spin moment of around $1.7\:\mu_B$ and an orbital moment of about $0.1\:\mu_B$, being in agreement with FeCo alloys and nanoclusters from previous works \cite{levzaic2007first,miranda2017dimensionality,gerasimov2021nature}. No clear trend is observed in spin and orbital moments with the average interatomic distance $d$. Nevertheless, an increase is noted with $n$, featuring the largest spin moment variation at $\sim1.8\%$ and the largest orbital moment variations at $\sim30-38\%$ for $x\geq40\%$, to be compared with $\Delta\alpha^{\textnormal{max}}$ --- i.e., giving somewhat a contribution but still not sufficient to explain the values shown in Table \ref{tab:effdampvar}. 

Therefore, in the scope of Eq. \ref{eq:kambersky}, the only quantity that remains to be analyzed is the total density of states (DOS) at the Fermi level. 
Significant correlation between $\alpha_i$ and $n({E_f})$ has been noted in this study and already in literature \cite{PhysRevB.94.214410}. It is shown in the second column of Fig.~\ref{fig:onsite_dos} that $n({E_f})$ exhibits a consistent enhancement trend with $d$ and $n$. The largest $\Delta n({E_f})$ found in each composition is shown in the bottom row of Table \ref{tab:effdampvar}, with qualitative agreement to the trends we observed for the effective damping. In detail, the local environment has a negligible impact on the spin-up channel's local DOS (LDOS), while it induces a shift in the spin-down channel towards lower energies, thus enhancing the corresponding LDOS at the Fermi level (data not shown).

Lastly, our findings reveal that the $\alpha_i$ is comparatively less sensitive on changes of $n$ for the Fe-centered clusters as well as on the local environment variation of the second nearest neighboring sites. Although the damping values of these clusters differ when compared to the pure VCA values, they are substantially lower than those observed in Co-centered cases and the cases varying in $d$ and $n$, respectively. For instance, when replacing the central Co atom with an Fe atom in Fe$_{53}$Co$_{47}$, and comparing the damping of clusters with different $n$, the maximum observed variation in damping – between the lowest and highest values – is $27\%$ (far less than in Co-centered cases). Additionally, our tests on several clusters with different second nearest neighbor environments show that the damping variations are negligible (less than $\sim5\%$). Furthermore, neither of these cases exhibit a clear trend in damping variation.
The underlying reason of this behavior is that the local environment impacts on $n({E_f})$ of Fe center clusters are negligible.

\subsection{Effective damping: towards the global frame}

\label{sec:global-perspective}

After analyzing the impact of specific Fe/Co configurations on Gilbert damping within a localized frame, we now discuss the EC-VCA results from a broader, more global perspective -- towards a random alloy model with short-range structure resolution. To this end, we proceed to calculate the average damping for each Co concentration.
Obviously, as detailed in Section \ref{subsec:local-effects-detailed}, as we move away from Co-poor alloys approaching Fe$_{50}$Co$_{50}$, the number of possible nonequivalent configurations greatly increases, which prevents a brute-force calculation of the precise average values for all $x$. As an illustration, we can briefly analyze the cases of $\rm Fe_{53}Co_{47}$ and $\rm Fe_{40}Co_{60}$. By utilizing a cluster expansion approach similar to that described in Refs. \cite{jiang2024influence, PhysRevB.81.245210, he2020atom}, 197 and 159 non-equivalent clusters are found for these compositions, respectively, without considering magnetic symmetry operations (SQA effect). Thus, a limited set has to be selected. We here choose to investigate one Fe-centered and one Co-centered cluster for each C$n$-$m$ group, including C0 and C1, for $0\%<x\leq40\%$; an example of this, applied to Co-centered clusters was already depicted in Fig. \ref{fig:cluster}. This approach yields a simplified and unbalanced average, which we argue is sufficiently representative to replicate the experimental trend.

\begin{figure}[!ht]
\centering
        \includegraphics[width=1\columnwidth]{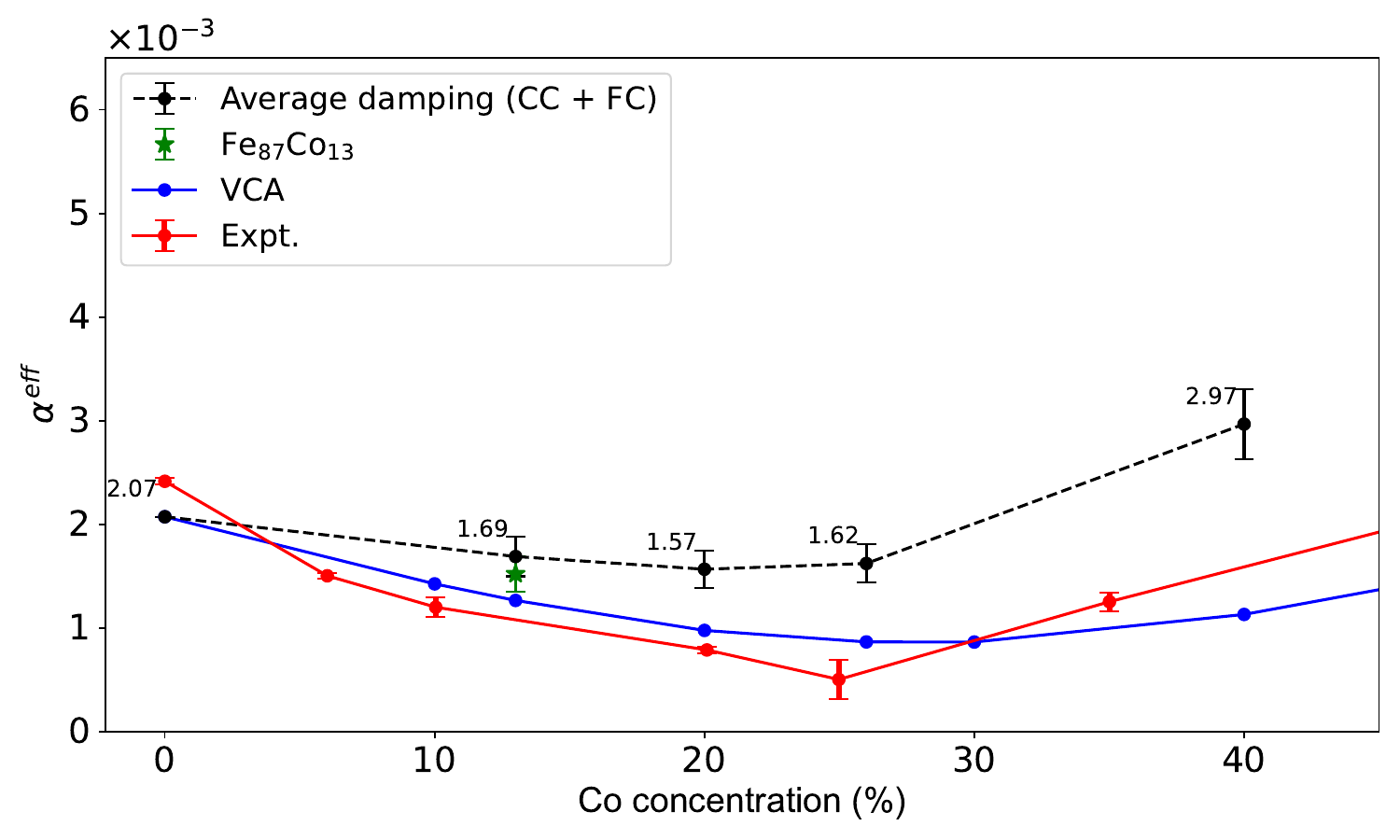}
        \caption{The effective intrinsic damping as a function of Co concentration. The averages (black dots) are performed over all considered Fe-centered and Co-centered clusters within each C$n$-$m$ group, and calculated in the FMR regime by Eq. \ref{eq:effective-alpha-parameter}. The blue and red dots show, respectively, the damping obtained from pure VCA calculations and experimental results from Ref. \cite{Schoen2016}. The average value for $\rm Fe_{87}Co_{13}$ obtained in Section \ref{subsec:local-effects-detailed} is depicted by the green star. Error bars are calculated as $\sim11\%$ of each value, as a result of the standard deviation obtained in the $\rm Fe_{87}Co_{13}$ case (see text). Lines are guides for the eyes.} 
        \label{fig:damping_comparison} 
\end{figure}

Figure \ref{fig:damping_comparison} shows the average effective damping, obtained by using Eq. \ref{eq:total-damping}, considering both the Co-centered and Fe-centered clusters among the different concentrations. On one hand, quantitatively we observe that the average damping shown in Fig. \ref{fig:damping_comparison} is enhanced in comparison with the experimental results and even with VCA calculations. This, however, is expected from the unbalanced average model we choose and has two simple explanations: (\textit{i}) the averages are taken from incomplete sets of clusters; and (\textit{ii}) the high-$\alpha_i$ clusters are included. This can be clearly seen by incorporating the complete set value obtained in Section \ref{subsec:local-effects-detailed} for $\rm Fe_{87}Co_{13}$, denoted by a green star in Figure \ref{fig:damping_comparison}, and comparing it with the average value obtained by our established selection criteria. As Figure \ref{fig:damping_comparison} shows, the average represented by the green star is rather close to the experimentally observed value.

On the other hand, qualitatively we note that the experimental trend (with a minimum around $x\sim20\%$) and order of magnitude are well reproduced by the calculated values shown by black dots, despite the limited set and high-$\alpha_i$ clusters included. The error bars follow the percentage ($\sim11\%$) found in Section \ref{subsec:local-effects-detailed}, although we expect increased standard deviations as $x\rightarrow50\%$, specially because outliers with considerable $\Delta\alpha^{\textnormal{eff}}$ become more common in those alloy concentrations. Therefore, although unusually high $\alpha_i$ clusters can emerge as a result of chemical disorder, those cases are hidden in real intrinsic damping measurements by statistical averages. This makes the local scenario described in Section \ref{subsec:damping variation} consistent with FMR experiments again.

\section{Conclusions and outlook}
\label{sec:conclusion}

In this study, we theoretically explored the role of local environments (or the short-range chemical disorder), specifically atomic configurations and compositions, on the intrinsic Gilbert damping. Our investigation is focused on various bcc FeCo alloys, namely Fe$_{100-x}$Co$_{x}$, with $x\in[0\%,60\%]$. With that perspective, a hybrid explicit/effective medium model (EC-VCA) was introduced. 
From the analysis of the complete set of configurations in the Fe$_{87}$Co$_{13}$ concentration, it was demonstrated that different spatial distributions of Fe/Co within the embedded cluster region can markedly affect the damping. Unlike other quantities such as the Heisenberg exchange interactions, damping is anisotropic with respect to the spin quantization axis, which further contributes (albeit slightly in FeCo) to the effects of chemical disorder. The computed average damping aligns well with both experiments and the pure VCA calculation. When low-temperature explicit atomistic spin dynamics simulations are performed, the influence of short-range disorder on local dynamics is observed to alter the overall relaxation rate in both nonlocal and effective damping formulations of the generalized Landau-Lifshitz-Gilbert (LLG) equation, even when such a small (15-atom) embedded region is considered.


Transitioning to configurations with higher Co content across the FeCo alloy series, for which inevitably only representative sets can be analyzed, the effective damping is found to be notably influenced by the atomic configuration at the nearest neighborhood. In the most extreme cases, a variation of about one order of magnitude is found. A direct correlation between effective damping and average the Co interatomic distance ($d$), as well as the quantity of Co atoms occupying the first neighborhood ($n$) were found. This local environment-dependent damping behavior was explained in light of the simplified Kambersky’s formula (Eq. \ref{eq:kambersky}), demonstrating a consistent correlation between damping and the local density of states at the Fermi level. In a global perspective (i.e., performing a configuration average), those differences in damping are masked by statistical averages, reconciling the local picture to what is typically observed in FMR experiments for FeCo.

Considering damping as a nonlocal (and inhomogeneous) quantity reveals several interesting characteristics, such as a pronounced dependence on the local environment, which are overlooked by effective medium theories. Such local environment-dependent damping may have implications, in real materials, on phenomena like the damping anisotropy observed in Fe$_{50}$Co$_{50}$, which was originally attributed to short-range configurational effects  \cite{PhysRevLett.122.117203}. Moreover, our results naturally encourage further exploration of the potential for local engeneering of damping, and the investigation into how defects affect both damping and the magnetization dynamics at local and global scales. 

\section{Acknowledgments}

Financial support from Vetenskapsrådet (grant numbers VR 2016-05980, VR 2019-05304, VR 2019-03666, and VR 2023-04239), and the Knut and Alice Wallenberg foundation (grant numbers 2018.0060, 2021.0246, 2022.0108, and 2022.0079) is acknowledged. Support from the Swedish Research Council (VR), the Swedish Energy Agency (Energimyndigheten), the European Research Council (854843-FASTCORR), eSSENCE and STandUP is acknowledged by O.E. Support from the Swedish Research Council (VR) is acknowledged by D.T., O.E.~and A.D. O.E. and A.D. acknowledge support from WISE, Wallenberg Initiative Materials Science, funded by the Knut and Alice Wallenberg Foundation. 
The China Scholarship Council (CSC) is acknowledged by Z.L. The computations/data handling were enabled by resources provided by the National Academic Infrastructure for Supercomputing in Sweden (NAISS), partially funded by the Swedish Research Council through grant agreement no. 2022-06725.
\bibliography{references.bib}
\bibliographystyle{apsrev4-2}

\clearpage
\appendix
\section{Number of Co atoms in each composition}
\label{sec:Cluster_configuration}

In Table \ref{tab:cluster} we enumerate the total number of Co atoms in the 15-atom cluster that corresponds to each alloy concentration ($N_{\textnormal{Co}}$), as well as the maximum/minimum number of Co atoms that can be placed within the nearest neighboring shell ($n_{\textnormal{max}}$ and $n_{\textnormal{min}}$ in their Co-centered configurations, respectively -- see Section \ref{subsec:clusters} for definitions).

\section{ASD simulations $J_{ij}$ parameters}
\label{sec:jijs-sets-asd}

\begin{figure}[!htb]
        \includegraphics[width=\columnwidth]{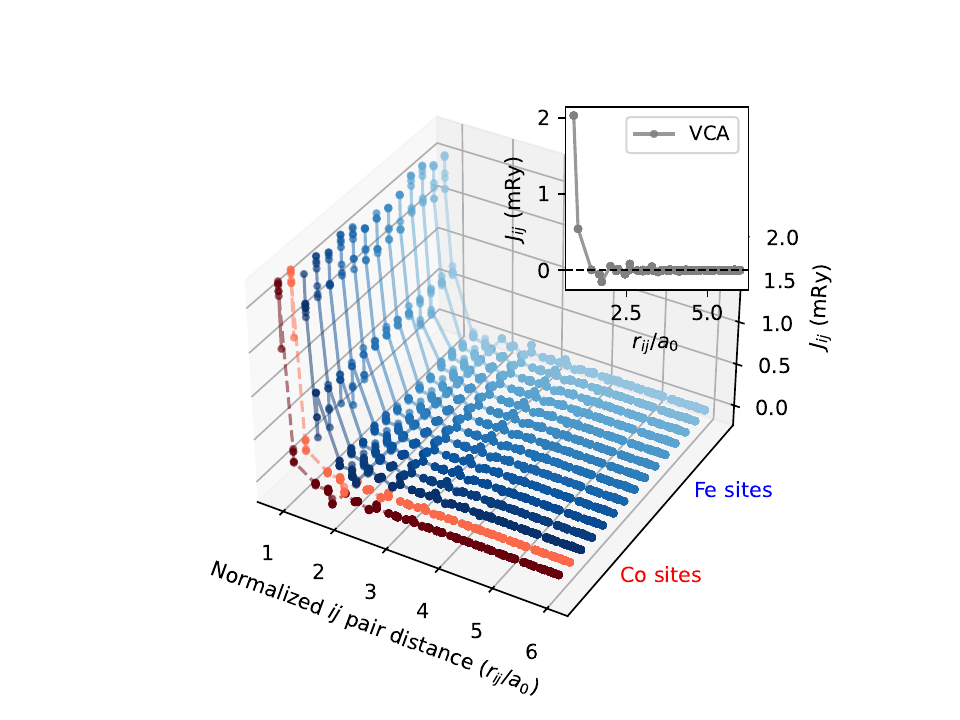} 
        \caption{Calculated Heisenberg exchange interaction parameters for each Fe (shades of blue) or Co (shades of red) atom within the embedded cluster illustrated in Fig. \ref{fig:cluster-example}, with each atom taken as a reference site.  The \textit{Inset} highlights, in gray, the $J_{ij}$ parameters computed for the VCA Fe$_{87}$Co$_{13}$. Lines are guides for the eyes.}
        \label{fig:embedded-cluster-jij} 
\end{figure}

Figure \ref{fig:embedded-cluster-jij} depicts the $J_{ij}$ parameters used to obtain the atomistic spin dynamics (ASD) results of Section \ref{sec:explicit-asd}. The plots correspond to the computed $J_{ij}$ values for each of the Fe and Co atoms of the embedded cluster of Fig. \ref{fig:cluster-example} taken as the reference site, as well as for the Fe$_{87}$Co$_{13}$ VCA (\textit{Inset}, in gray).

\begin{table*}[htb]
\Large
\caption{\label{tab:cluster}
The total number of Co atoms ($N_{\textnormal{Co}}$) and the maximum (minimum) amount of Co in the nearest-neighborhood $n_{\textnormal{max}}$ ($n_{\textnormal{min}}$) within the cluster and in each composition. All these clusters refer to Co-centered configurations.}

\begin{ruledtabular}
\resizebox{0.5\textwidth}{!}{
\begin{tabular}{cccccc}
\multicolumn{1}{c}{}&
\multicolumn{1}{c}{Composition} & {$N_{\textnormal{Co}}$} & $n_{\textnormal{max}}$ & $n_{\textnormal{min}}$&\\
\colrule \\
{}&Fe & 0 & 0 &0\\
{}&Fe$_{87}$Co$_{13}$ & 2 & 1&0\\ 
{}&Fe$_{80}$Co$_{20}$ & 3 & 2&0\\ 
{}&Fe$_{74}$Co$_{26}$ & 4 & 3&0\\ 
{}&Fe$_{60}$Co$_{40}$ & 6 & 5&0\\ 
{}&Fe$_{53}$Co$_{47}$ & 7 & 6&0\\ 
{}&Fe$_{40}$Co$_{60}$ & 9 & 8&2\\
\end{tabular}
}
\end{ruledtabular}
\end{table*}

\section{Convergence of off-diagonal damping terms}
\label{sec:off-diagonal-convergence}

\begin{figure}[!htb]
        \includegraphics[width=\columnwidth]{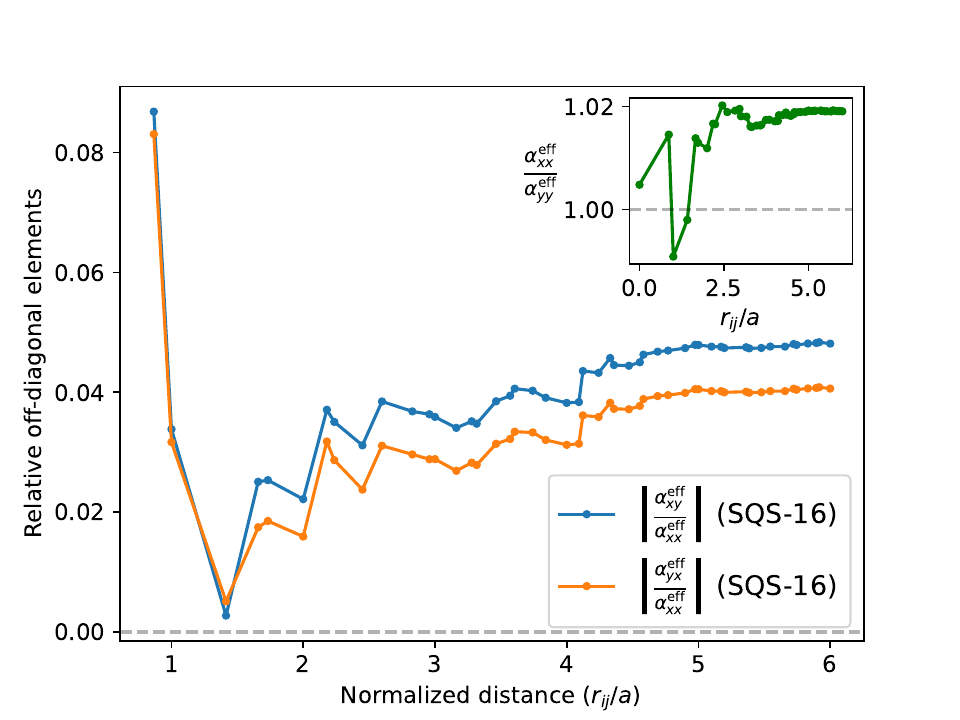} 
        \caption{Relative values of the off-diagonal terms $\alpha^{\textnormal{eff}}_{xy}$ (blue dots) and $\alpha^{\textnormal{eff}}_{yx}$ (orange dots) in the damping tensor $\hat{\alpha}^{\textnormal{eff}}$ for a SQS-16 cell of bcc Fe$_{50}$Co$_{50}$. The values were obtained using the formulation of Eq. \ref{eq:effective-tensor-elements}. \textit{Inset}: asymmetry between the diagonal terms $\alpha^{\textnormal{eff}}_{xx}$ and $\alpha^{\textnormal{eff}}_{yy}$, defined as $\frac{\alpha^{\textnormal{eff}}_{xx}}{\alpha^{\textnormal{eff}}_{yy}}$. The lines are guides for the eyes.}
        \label{fig:off-diag-sqs} 
\end{figure}

Figure \ref{fig:off-diag-sqs} shows the comparison of $\alpha^{\textnormal{eff}}_{xy}$ and $\alpha^{\textnormal{eff}}_{yx}$ with the diagonal terms, as well as their asymmetry $\left(\frac{\alpha^{\textnormal{eff}}_{xx}}{\alpha^{\textnormal{eff}}_{yy}}\right)$. 

\section{VCA approach in FeCo systems}\label{VCA}
\begin{figure}[!htb]
        \includegraphics[width=\columnwidth]{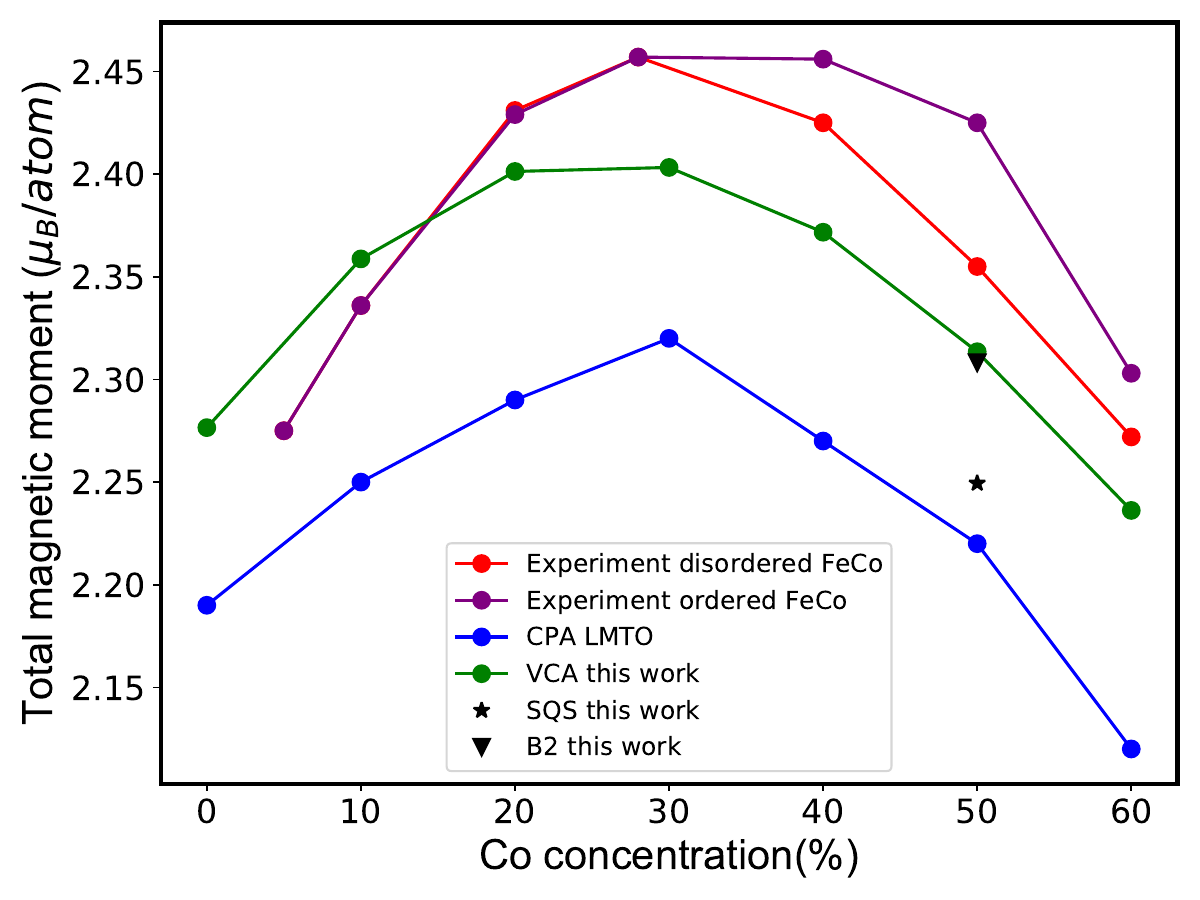} 
        \caption{Total magnetic moments (spin + orbital) as a function of the Co concentration, obtained for the FeCo alloys, exhibiting the well-known Slater-Pauling behavior. The calculations in this work are performed using 3 different models: VCA (green circles), SQS-16 (black star), and ordered B2 structure (black triangle). The blue dots represent the data obtained by using the CPA-LMTO method (extracted from Ref. \cite{PhysRevB.59.419}). Experimental data from Ref. \cite{Bardos1969} are depicted by the red and purple circles, representing the measurements for disordered and ordered FeCo alloys, respectively. Lines are guides for the eyes.} 
       
        \label{fig:Slater-Pauling} 
\end{figure}

In the proposed EC-VCA model, the VCA was used to describe the external (non-explicit) medium. Despite its simple nature, the VCA approach has been demonstrated to be a reasonably valid method for investigating the properties of Fe$_x$Co$_{1-x}$ alloys. This has been discussed in the context of: (\textit{i}) magnetic moments (both spin and orbital) as corroborated by the replication of the well-known Slater-Pauling curve (see Fig. \ref{fig:Slater-Pauling}) \cite{PhysRevB.45.12911,PhysRevB.74.104422}; (\textit{ii}) transport properties \cite{PhysRevB.88.104408}; (\textit{iii}) a \textit{qualitatively} correct behavior of the magnetic anisotropy energy, as discussed in Refs. \cite{PhysRevB.89.144403,Neise2011,PhysRevB.86.174430,PhysRevB.93.224425}; and (\textit{iv}) a \textit{qualitatively} accurate trend of the total intrinsic Gilbert damping, as shown in Ref. \cite{PhysRevB.108.014433}, for which VCA is able to capture the remarkable low value at $\sim20-30\%$ of Co concentration, physically attributed to a pronounced minimum in the density of states at the Fermi level \cite{Schoen2016}. Although a non-negligible disorder is present in the spin-minority channel of bcc FeCo alloys \cite{PhysRevB.86.174430,PhysRevB.49.3352}, in an effective picture (in the context of Eq. \ref{eq:kambersky}), the average density of states of an explicit supercell is well reproduced by VCA \cite{PhysRevB.108.014433}. This trend, initially obtained via CPA analysis \cite{PhysRevLett.107.066603,PhysRevB.87.014430,PhysRevB.92.214407}, was later confirmed by experiments \cite{Schoen2016}.

 To give a more comprehensive relation between VCA, CPA and experimental results, we present a comparison of the Curie temperature ($\rm T_C$), spin wave stiffness ($\rm D_{ex}$) and Gilbert damping of Fe and Fe$_{1-x}$Co$_{x}$ alloys in Tables \ref{table1}-\ref{table3}. We found that the $T_C$, $\rm D_{ex}$ and dampings obtained from VCA and CPA show a good agreement, with acceptable discrepancies comparing to experiments.
\begin{table}[h!]
\caption{\label{table1}
The Curie temperature ($\rm T_C$) of the pure bcc Fe and Fe$_{1-x}$Co$_{x}$ alloys. In this work, whenever $x>0$, the electronic structure (and derived magnetic parameters) are calculated considering the VCA approach. The Curie temperature is evaluated from mean-field approximation ($\rm T_{C}^{MFA}$), random phase approximation  ($\rm T_{C}^{RPA}$), and Monte Carlo simulations  ($\rm T_{C}^{MC}$). For comparison, the theoretical and experimental value of $\rm T_C$ is shown. In other theoretical works, the CPA approach is adopted. } 
\resizebox{\columnwidth}{!}{
\begin{tabular}{ccccccc}

\hline
\hline
\multicolumn{4}{c}{This work}&\multicolumn{2}{c}{Other works}\\
\hline
\multicolumn{1}{c}{}&
\multicolumn{1}{c}{$\rm T_{C}^{MFA}$(K)}& $\rm T_{C}^{RPA}$(K) &$\rm T_C^{MC}$(K)& Expt. & Theory\\
\hline
bcc Fe  & 1274 & 919 & 960 &1044\cite{campbell1982ferromagnetic}  & 900$-$1238\cite{PhysRevLett.77.334,PhysRevB.55.14975,PhysRevB.75.054402,levzaic2007first,PhysRevB.95.184432}\\

bcc Fe$_{90}$Co$_{10}$  & 1712 & 1086 & 1300 &1164\cite{nishizawa1984co}& 1069\cite{PhysRevB.104.024403}\\

bcc Fe$_{80}$Co$_{20}$  & 2064 & 1426 & 1570 & 1225\cite{nishizawa1984co} & 1369\cite{PhysRevB.104.024403}\\

bcc Fe$_{70}$Co$_{30}$  & 2256 & 1667 & 1740  & $\sim1260$\cite{nishizawa1984co,karipoth2013synthesis}& 1490$-$1656\cite{levzaic2007first,PhysRevB.95.184432,PhysRevB.104.024403} \\

bcc Fe$_{60}$Co$_{40}$  & 2304 & 1790 & 1800 & 1268\cite{nishizawa1984co}& 1547\cite{PhysRevB.104.024403}\\

bcc Fe$_{50}$Co$_{50}$  & 2230 & 1782 & 1780 &  1253$–$1370\cite{nishizawa1984co,kawahara2003high} &1568$-$1634\cite{PhysRevB.104.024403,levzaic2007first,PhysRevB.95.184432} \\

bcc Fe$_{40}$Co$_{60}$  & 2067 & 1670 & 960& 1211\cite{karipoth2013synthesis}& $-$\\
\hline
\hline

\end{tabular}

}
\end{table}

\begin{table}[h!]

\caption{\label{table2}
The comparison of exchange stiffness ($\rm D_{ex}$) in this work and other works for the pure bcc Fe and Fe$_{1-x}$Co$_{x}$ alloys. } 
\resizebox{\columnwidth}{!}{
\begin{tabular}{cccc}
\hline
\hline
\multicolumn{1}{c}{}&This work&\multicolumn{2}{c}{Other works}\\
\hline
\multicolumn{1}{c}{}&
$\rm D_{ex}$ (meV$\cdot$\AA$^{2}$) & Expt. & Theory \\
\hline \\
bcc Fe  &  290 & 266$-$314\cite{stringfellow1968observation,PhysRevB.7.336,PhysRevLett.15.146} &247$-$410\cite{PhysRevB.55.14975,PhysRevB.104.024403}\\

bcc Fe$_{90}$Co$_{10}$  & 330 & 
$\sim345$ \cite{PhysRevLett.14.698} &
405\cite{PhysRevB.104.024403}\\

bcc Fe$_{80}$Co$_{20}$  & 411 & 
$\sim360$ \cite{PhysRevLett.14.698}
& 477\cite{PhysRevB.104.024403}\\

bcc Fe$_{70}$Co$_{30}$  & 493 & $\sim410$$-$470\cite{liu1994exchange,PhysRevLett.14.698} & 567\cite{PhysRevB.104.024403}\\

bcc Fe$_{60}$Co$_{40}$  & 561 &$\sim445$$-$530\cite{liu1994exchange,PhysRevLett.14.698} & 624\cite{PhysRevB.104.024403}\\

bcc Fe$_{50}$Co$_{50}$  & 592 &800\cite{liu1994exchange} &526$-$677\cite{PhysRevB.95.184432,PhysRevB.104.024403} \\

bcc Fe$_{40}$Co$_{60}$  & 578 
& $\sim480$ \cite{PhysRevLett.14.698}
&466\cite{PhysRevB.95.184432}\\
\hline
\hline
\end{tabular}
}
\end{table}

\begin{table}[h!]
\caption{\label{table3}
 The comparison of damping ($\alpha$) in this work and other works for the pure bcc Fe and Fe$_{1-x}$Co$_{x}$ alloys. } 
 \resizebox{\columnwidth}{!}{
 \begin{tabular}{cccc}
\hline
\hline

\multicolumn{1}{c}{}&This work&\multicolumn{2}{c}{Other works}\\
\hline
\multicolumn{1}{c}{}&
$\rm \alpha$ ($\times10^{-3}$) & Expt. & Theory \\
\hline \\
bcc Fe  &  2.1 &1.9$-$7.2\cite{Schoen2016,Oogane2006,PhysRevB.87.014430,PhysRevLett.124.157201, PhysRevLett.98.117601,PhysRevB.10.179,PhysRevB.18.4856,PhysRevB.95.134411}&1.3$-$3.2\cite{PhysRevB.87.014430,PhysRevLett.99.027204,Hiramatsu2021} \\

bcc Fe$_{90}$Co$_{10}$  & 1.4 &1.2\cite{Schoen2016} &0.5$-$0.8\cite{PhysRevB.92.214407,PhysRevB.87.014430, PhysRevB.98.104406} \\

bcc Fe$_{80}$Co$_{20}$  & 1.0 &  0.8\cite{Schoen2016}&0.4$-$0.6\cite{PhysRevB.92.214407,PhysRevB.87.014430,PhysRevB.98.104406}\\

bcc Fe$_{70}$Co$_{30}$ & 0.9 & 0.5$-$1.7\cite{Schoen2016, Oogane2006,Lee2017}&0.5$-$0.8\cite{PhysRevB.92.214407,PhysRevB.87.014430} \\

bcc Fe$_{60}$Co$_{40}$  & 1.1  & $\sim1.1$\footnote{bcc Fe$_{65}$Co$_{35}$ \cite{PhysRevB.95.134411}} & 0.8$-$1.1\cite{PhysRevB.92.214407,PhysRevB.87.014430,PhysRevB.98.104406}\\

bcc Fe$_{50}$Co$_{50}$  & 1.6 & 2.0$-$3.2\cite{Schoen2016, Oogane2006,PhysRevLett.122.117203} & 1.0$-$1.3\cite{PhysRevB.92.214407,PhysRevB.87.014430,PhysRevB.98.104406}\\

bcc Fe$_{40}$Co$_{60}$  & 1.6 &1.3$-$2.5\cite{Schoen2016}&1.6\cite{PhysRevB.92.214407,PhysRevB.87.014430} \\
\hline
\hline
\end{tabular}

}
\end{table}

\section{ Effective damping  }
\label{sec:damping_formula}
As mentioned in Section \ref{sec:methods}, in the FMR experiments the magnetic moments are excited in a coherent mode. Thus, a macrospin description with the summed total effective magnetization $\vec{M}^{\textnormal{eff}}_i$ reads

\begin{equation}\label{eq:sum-damping}
\begin{split}
\vec{M}^{\textnormal{eff}}_i =\sum_i \vec{M}_i.
\end{split}
\end{equation}

The dynamics of this macrospin is well described by the Landau-Lifshitz-Gilbert equation 

\begin{equation}\label{eq:llg-meff}
\frac{\partial \vec{M}^{\textnormal{eff}}}{\partial t}=\vec{M^{\textnormal{eff}}}\times\left(-\gamma\vec{B}+\frac{\alpha^{\textnormal{eff}}}{\left|\vec{M}^{\textnormal{eff}}\right|} \frac{\partial \vec{M}^{\textnormal{eff}}}{\partial t}\right),
\end{equation}

\noindent where $\gamma$ is the gyromagnetic ratio, $\vec{B}$ is the effective field acting on the macrospin and $\left|\vec{M}^{\textnormal{eff}}\right|=\left|\sum_i\vec{m}_i\right|\leq\sum_i\left|\vec{m}_{i}\right|$. Since we aim to describe a coherent motion of the individual spins, i.e., a ferromagnetic collinear state at all times $t$, by construction we have $\left|\vec{M}^{\textnormal{eff}}\right|=\sum_i\left|\vec{m}_{i}\right|$  and $\vec{B}$ isotropic with respect to all atomic sites $i$ ($\vec{B}_i=\vec{B}$). A general form of LLG equation incorporating the non-local damping (first assuming it to be a scalar parameter) is:

\begin{equation}\label{eq:llg_nonlocal}
\frac{\partial \vec{m}_i}{\partial t}=\vec{m}_i\times\left(-\gamma \left[\vec{B}_i+\vec{b}_i(t)\right]+\sum_j\frac{\alpha_{ij}}{m_j}\frac{\partial \vec{m}_j}{\partial t}\right).
\end{equation}

Working with Eq. \ref{eq:llg-meff} we have:

\begin{align}\label{eq:meff-deduction}
\frac{\partial \vec{M}^{\textnormal{eff}}}{\partial t}
&=\sum_i \frac{\partial \vec{m}_i}{\partial t} \nonumber\\
&=\sum_i\left[\vec{m}_i\times(-\gamma \vec{B}_i+\sum_j\frac{\alpha_{ij}}{m_j} \frac{\partial \vec{m}_j}{\partial t} )\right] \nonumber\\
&=-\gamma \left[\sum_i\vec{m}_i\right]\times \vec{B}+\left[\sum_i{\vec{m}_i}\right]\times \left[\sum_j\frac{\alpha_{ij}}{m_j}\frac{\partial \vec{m}_j}{\partial t}\right].
\end{align}

Therefore, in the FMR regime, one can compare Eqs. \ref{eq:llg-meff} and \ref{eq:meff-deduction}, resulting in 

\begin{equation}\label{eq:comparison}
\vec{M}^{\textnormal{eff}} \times \frac{\alpha^{\textnormal{eff}}}{\left|\vec{M}^{\textnormal{eff}}\right|}\frac{\partial \vec{M}^{\textnormal{eff}}}{\partial t}  = \sum_i\vec{m}_i\times\sum_j \frac{\alpha_{ij}}{m_j} \frac{\partial \vec{m}_j}{\partial t} . 
\end{equation}

Let us write $\vec{M}^{\textnormal{eff}} = \left|\vec{M}^{\textnormal{eff}}\right|\vec{e}^{\textnormal{eff}} $ and $\vec{m}_i = \left|\vec{m}_i\right|\vec{e}_i$, where $\vec{e}^{\textnormal{eff}}$ and $\vec{e}_i$ are the spin moment orientation of the macrospin and the atomic spin at site $i$, respectively, at a given time $t$. Since we describe a coherent precession, then it is straightforward that $\frac{\partial \vec{e}_i}{\partial t}=\frac{\partial \vec{e}^{\textnormal{eff}}}{\partial t}$ and $\vec{e}_i=\vec{e}^{\textnormal{eff}}$. Hence, by Eq. \ref{eq:comparison}, we obtain 

\begin{align}\label{eq:effective-alpha-parameter}
\alpha^{\textnormal{eff}}&=\frac{1}{M^{\textnormal{eff}}}\sum_{ij} \left|\vec{m}_i \right|\alpha_{ij} \nonumber \\ 
&= \frac{1}{\sum_i \left|\vec{m}_i \right| } \sum_{ij} \left|\vec{m}_i \right|\alpha_{ij} = \frac{1}{\sum_i \left|\vec{m}_i \right| } \sum_{i}\left|\vec{m}_i \right|\alpha_i.
\end{align}

Here, we define the effective damping of spin moment $i$ as $\alpha_i=\sum_j{\alpha_{ij}}$ \cite{PhysRevB.108.014433}. For the general case where the Gilbert damping is expressed as a tensor $\boldsymbol{\alpha}_{ij}$, it becomes harder to disentangle a closed form for $\alpha^{\textnormal{eff}}$, and the effective damping parameter should be extracted directly from simulations containing the explicit $\boldsymbol{\alpha}_{ij}$ expression. However, if we assume that the macrospin is also subjected to an effective damping in the tensor form, as suggested by previous works (e.g., Ref. \cite{PhysRevB.72.064450,PhysRevB.84.172403}), so that the term $\frac{\alpha^{\textnormal{eff}}}{\left|\vec{M}^{\textnormal{eff}}\right|}\frac{\partial \vec{M}^{\textnormal{eff}}}{\partial t}$ is replaced by $\frac{\boldsymbol{\alpha}^{\textnormal{eff}}}{\left|\vec{M}^{\textnormal{eff}}\right|}\cdot\frac{\partial \vec{M}^{\textnormal{eff}}}{\partial t}$, then for each element of $\boldsymbol{\alpha}^{\textnormal{eff}}$ we have

\begin{equation}
\label{eq:effective-tensor-elements}
\alpha^{\textnormal{eff}}_{\mu\nu} = \frac{1}{M^{\textnormal{eff}}}\sum_{ij} \left|\vec{m}_i \right|\alpha_{ij}^{\mu\nu},
\end{equation}

\noindent where $\mu,\nu\in\{x,y,z\}$. This result can be derived by applying the same reasoning of Eqs. \ref{eq:meff-deduction}-\ref{eq:effective-alpha-parameter}, however considering that both effective and atomistic LLG equations assume the general form of a tensorial Gilbert damping ($\boldsymbol{\alpha}^{\textnormal{eff}}$ and $\boldsymbol{\alpha}_{ij}$).

\end{document}